\documentclass{pasj02}
\usepackage{natbib}
\usepackage[switch,mathlines]{lineno}
\usepackage{url}
\usepackage{graphicx}	
\usepackage{amsmath}	
\usepackage{tikz}
\usepackage{longtable}
\usepackage{multirow}
\usepackage{threeparttable}

\usepackage{natbib}
\usepackage[breaklinks,colorlinks,
   urlcolor=blue,citecolor=blue,linkcolor=blue]{hyperref}
\renewcommand{\refname}{\normalsize References}
\usepackage{lipsum}
\let\OLDthebibliography\thebibliography
\renewcommand\thebibliography[1]{
  \OLDthebibliography{#1}
  \setlength{\parskip}{0pt}
  \setlength{\itemsep}{0pt plus 0.3ex}
}

\Received{$\langle$26-Apr-2026$\rangle$}
\Accepted{$\langle$26-Jun-2026$\rangle$}
\Published{$\langle$publication date$\rangle$}

\definecolor{lime}{HTML}{A6CE39}
\DeclareRobustCommand{\orcidicon}{
	\begin{tikzpicture}
	\draw[lime, fill=lime] (0,0) 
	circle [radius=0.13] 
	node[white] {{\fontfamily{qag}\selectfont \tiny ID}};
	\draw[white, fill=white] (-0.0625,0.095) 
	circle [radius=0.007];
	\end{tikzpicture}
	\hspace{-2mm}
}
\foreach \x in {A, ..., Z}{\expandafter\xdef\csname orcid\x\endcsname{\noexpand\href{https://orcid.org/\csname orcidauthor\x\endcsname}
			{\noexpand\orcidicon}}
}

\begin{document}

\title{Fe K$\alpha$ equivalent-width mapping with 3D radiative transfer calculation: A general model and application to the RS Canum Venaticorum–type stars with XRISM/Resolve}

\author{Shun Inoue\altaffilmark{1,$*$\orcidA}} 
\author{Masahiro Tsujimoto\altaffilmark{2,3\orcidF}}
\author{Takayuki Hayashi\altaffilmark{1\orcidG}}
\author{Teruaki Enoto\altaffilmark{1,4\orcidB}}
\author{Yuta Notsu\altaffilmark{5,6\orcidC}} 
\author{Hiroyuki Uchida\altaffilmark{1\orcidD}}
\author{Miki Kurihara\altaffilmark{2,3\orcidE}}

\altaffiltext{1}{Department of Physics, Kyoto University, Kitashirakawa-Oiwake-cho, Sakyo-ku, Kyoto 606-8502, Japan}
\altaffiltext{2}{Institute of Space and Astronautical Science (ISAS), Japan Aerospace Exploration Agency (JAXA), 3-1-1 Yoshino-dai, Chuo-ku,
Sagamihara, Kanagawa 252-5210, Japan}
\altaffiltext{3}{Department of Astronomy, Graduate School of Science, The University of Tokyo, 7-3-1 Hongo, Bunkyo-ku, Tokyo 113-0033, Japan}
\altaffiltext{4}{RIKEN Cluster for Pioneering Research, 2-1 Hirosawa, Wako, Saitama 351-0198, Japan}
\altaffiltext{5}{Laboratory for Atmospheric and Space Physics, University of Colorado Boulder, 3665 Discovery Drive, Boulder, CO 80303, USA}
\altaffiltext{6}{National Solar Observatory, 3665 Discovery Drive, Boulder, CO 80303, USA}
\email{inoue@cr.scphys.kyoto-u.ac.jp}
 
\KeyWords{radiative transfer --- X-rays: general --- X-rays: stars --- stars: coronae --- stars: late-type --- }

\maketitle

\begin{abstract}

The Fe~K$\alpha$ fluorescence line at 6.4 keV has long been used to probe the relative geometry between photoionizing X-ray sources and surrounding cold material in a wide range of astrophysical systems.
With the advent of the X-ray microcalorimeter XRISM/Resolve, Fe~K$\alpha$ lines with equivalent widths down to $\sim 5$~eV—previously inaccessible—are now detectable, and even non-detections can place upper limits of a few eV, making non-detections themselves valuable for constraining the geometry.
Considering that Fe~K$\alpha$-based geometric diagnostics are entering a new stage in the microcalorimeter era, we present Fe~K$\alpha$ equivalent-width maps computed with the three-dimensional Monte Carlo radiative-transfer code \texttt{SKIRT} for a generalized configuration consisting of a spherical reflector of radius $R_{*}$ and a point source located at a height $h$ above the surface.
The equivalent-width maps exhibit two characteristic features: (1) an increase toward the center of the projected surface of the sphere;
and (2) an overall decrease with increasing $h/R_{*}$.
The key point is that we confirm these features for equivalent widths of $< 40$~eV, a regime that has become accessible for the first time thanks to the improved detection threshold from $\sim 50$~eV with Chandra/HETG to $\sim 5$~eV with XRISM/Resolve.
As an illustrative application, we compare the maps with XRISM/Resolve spectra of three RS~Canum Venaticorum-type stars (GT~Muscae, $\sigma$~Geminorum, and HR~1099) and constrain the locations of the flare loop and coronal bright points in these systems.
Because the maps are constructed for a highly generalized point-source--spherical-reflector geometry, they are readily applicable to many other objects, including X-ray binaries and cataclysmic variables.

\end{abstract}


\begin{table*}[]
\centering
\caption{Resolve data set of the RS CVn-type stars used in this study. } 
\label{tab:best-fit}
\renewcommand{\arraystretch}{1.25}
\begin{threeparttable}
\begin{tabular}{ccccccc}
\hline
Star                     & Sequence number & Obs. start time & Exposure$^{*}$ & Phase & CR$^{\dagger}$ & EW$_{\mathrm{K\alpha}}^{\ddagger}$ \\
                         &       & (UT) & (ks) & & (s$^{-1}$) & (eV) \\ \hline
GT Mus                   & 300000010 & 2024-08-13T02:03:39 & 199 & Quiescence & 0.34 & $23.8^{+5.7}_{-8.3}$ \\ \hline
$\sigma$ Gem             & 202063010 & 2025-11-06T22:19:04 & 109 & Quiescence & 0.33 & $< 3.9$ \\ \hline
\multirow{2}{*}{HR 1099} & 100011011 & 2024-03-07T13:00:03 & 51 & Flare & 0.75 & $4.6^{+3.7}_{-2.9}$ \\
                         & 100011010--100011012 & 2024-03-06T01:21:54 & 180 & Quiescence & 0.41 & $< 3.6$ \\ \hline
\end{tabular}
\begin{tablenotes}
\footnotesize
\item[$^{*}$] Effective exposure after the standard screening.
\item[$^{\dagger}$] Count rate of Resolve after the standard screening in all energy bands.
\item[$^{\ddagger}$] Equivalent width of the sum of the Fe K$\alpha_{1}$ and Fe K$\alpha_{2}$ lines.
\end{tablenotes}
\end{threeparttable}
\end{table*}

\section{Introduction}
The Fe~K$\alpha$ line at 6.4 keV ($2p \rightarrow 1s$) is a prominent feature in astrophysical X-ray spectra due to iron's high fluorescence yield and its relatively high abundance in the Universe.
This line is produced via the inner-shell ionization of neutral or low-ionized iron when cool material is irradiated by X-ray photons or electrons with energies above the Fe K-edge (7.11~keV), creating K-shell vacancies that are filled by radiative transitions.
Such a configuration---a hot X-ray source illuminating cooler matter that contains neutral or low-ionized iron---is common across astrophysical environments, including active galactic nuclei, cataclysmic variables, X-ray binaries, and stars.
Because the observed Fe~K$\alpha$ line equivalent-width depends on the relative geometry among the observer, the hot X-ray source, and the cool reflector, it has been widely used as a geometric diagnostic in various systems. 
Also, in some objects with strong gravitational fields, the Fe~K$\alpha$ line can be broadened up to several keV by relativistic effects, and its line profile has been used to infer the geometry \citep[e.g.,][]{Tanaka_1995, Miller_2002, Liu_2015, Mantovani_2016}. In this paper, however, we focus exclusively on the narrow Fe~K$\alpha$ line.

Modeling of the Fe~K$\alpha$ line began in the 1980s, when the Fe~K$\alpha$ emission from Galactic X-ray binaries and Seyfert galaxies started to be detected \citep[e.g.,][]{Ohashi_1984, Ohashi_1984b, Matsuoka_1986, Nagase_1986, Makishima_1987, Koyama_1989, Kunieda_1990}.
For example, \citet{Inoue_1985} performed Monte Carlo simulations for a configuration consisting of an X-ray source surrounded by spherical cold gas and presented the Fe~K$\alpha$ equivalent width as a function of the gas column density. 
\citet{Krolik_1987} combined the radiative transfer code of \citet{Krolik_1984} with an analytic approach to predict the Fe~K$\alpha$ equivalent width expected from the torus of Seyfert galaxies.
\citet{George_1991}, one of the most widely cited works on Fe~K$\alpha$ line modeling, quantified with Monte Carlo radiative-transfer calculations how the Fe~K$\alpha$ equivalent width depends on the observer's inclination in a system where X-rays from a point source illuminate a flat, circular slab.
The calculation setup adopted by \citet{George_1991} is sufficiently general to be applicable to both X-ray binaries and AGN.

Thirty-five years have passed since their pioneering work, and two major developments have reshaped the landscape.
First, advances in computing have greatly expanded the flexibility of radiative-transfer modeling, allowing one to specify geometries, radiation sources, properties of the transfer medium, spatial grids, and detectors with high freedom.
Second, the X-ray microcalorimeter Resolve \citep{Kelley_2025, Ishisaki_2025} onboard XRISM \citep{Tashiro_2025} has ushered in high-resolution spectroscopy with an energy resolving power of $R \equiv E/\Delta E = 1400$ \citep{Porter_2025} and an effective area of $A=174~\mathrm{cm^2}$ at the Fe~K$\alpha$ energy \citep{Ishisaki_2022, Ishisaki_2025}.
Before the advent of Resolve, Chandra/HETG \citep{Canizares_2005} was arguably the most suitable instrument for detecting Fe~K$\alpha$, with an energy resolving power of $R=167$ and an effective area of $A=20~\mathrm{cm^{2}}$ at 6.4~keV.
Given that HETG could reach a minimum detectable equivalent width of $\sim 50$~eV \citep[e.g.,][]{Torrejon_2010}, the improvement in the figure of merit for the line detection $\sqrt{A/\Delta E}$ implies that Resolve should be able to detect Fe~K$\alpha$ lines that are $\sim 8.5$ times weaker, i.e., with equivalent widths down to $\sim 5$~eV.
Moreover, even when the line is not detected, Resolve can place upper limits of a few eV on the equivalent width, yielding constraints more than an order of magnitude tighter than those from previous instruments.
These improved detection and upper limits provided by Resolve will enable Fe~K$\alpha$-based geometric constraints for a much broader range of objects than before, regardless of whether the line is detected or not.

\begin{figure*}[t]
\begin{center}
\includegraphics[width=0.98\textwidth]{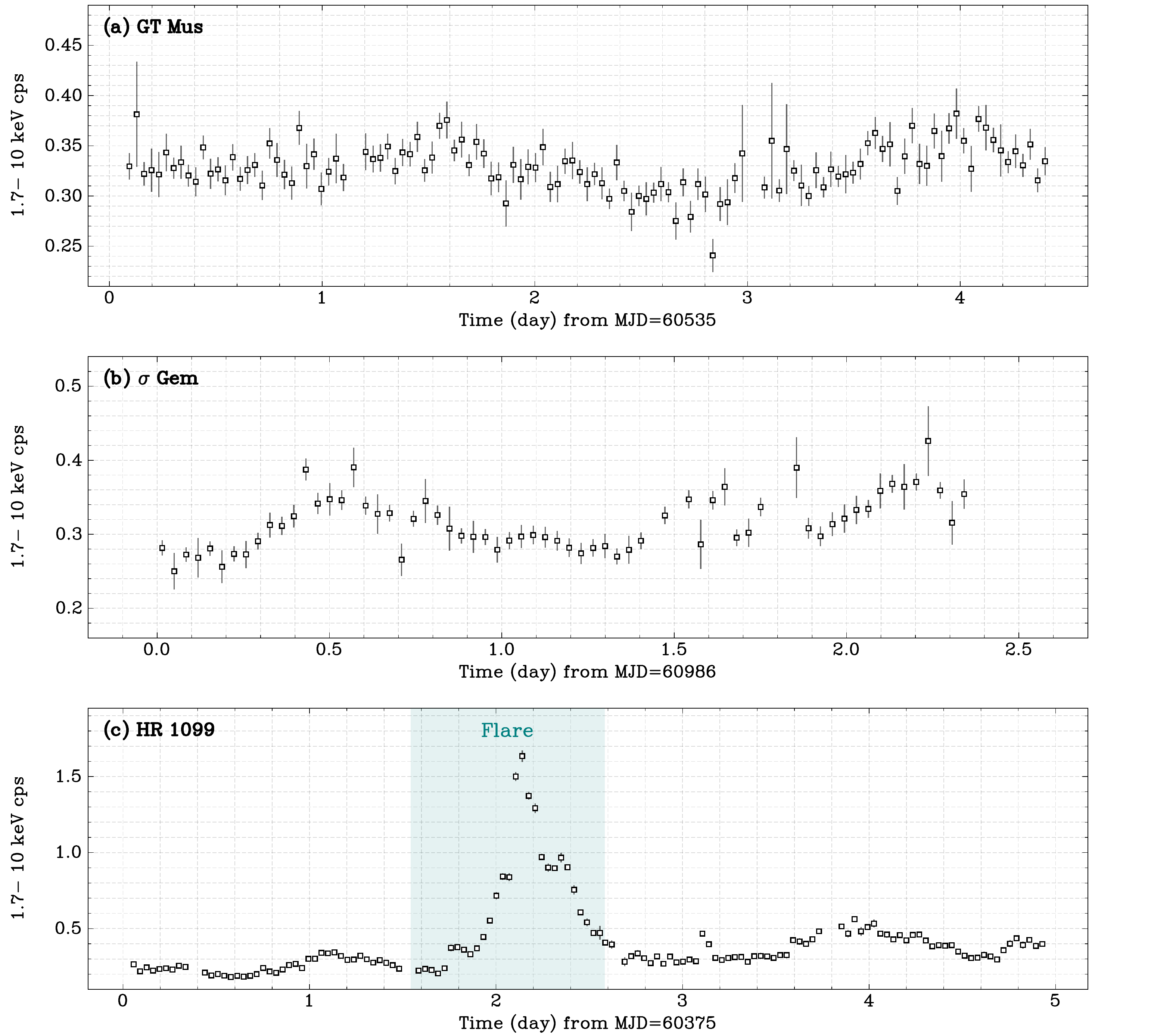}
\caption{Resolve light curve at 1.7--10.0 keV after the standard screening for (a) GT Mus, (b) $\sigma$ Gem, and (c) HR 1099. Time binsize is 3 ks for all curves. The green shaded area in panel c shows the time defined as the flare phase in \citet{Kurihara_2025b}. 
{Alt text: Three line graphs of the light curves with x-axis showing MJD and y-axis showing count rate. }
}
\label{fig:GTMus_lc}
 \end{center}
\end{figure*}

\begin{figure*}[]
\begin{center}
\includegraphics[width=0.98\textwidth]{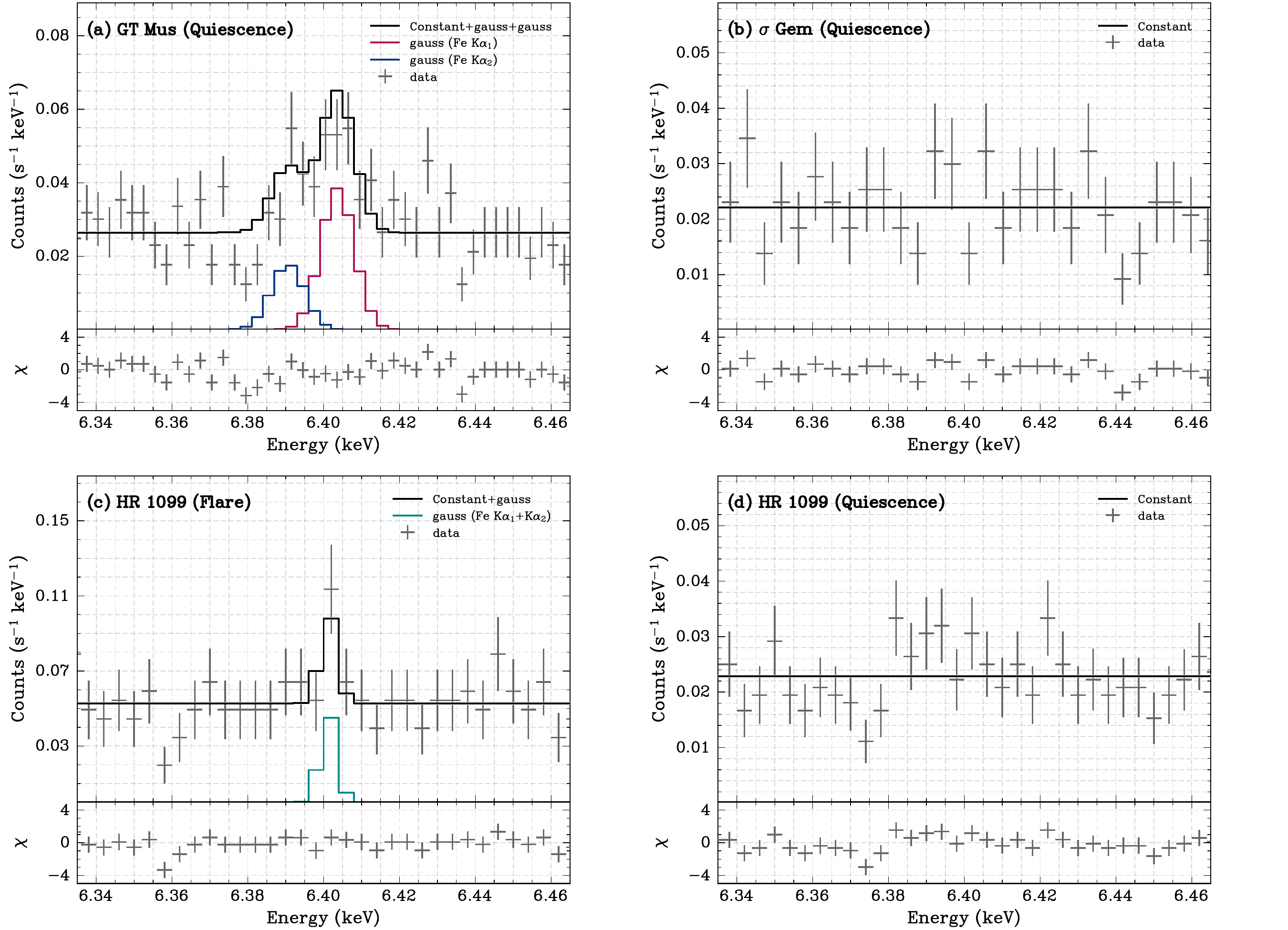}
\caption{Narrow-band Resolve spectral analysis around Fe K$\alpha$ line for (a) GT Mus, (b) $\sigma$ Gem, (c) HR 1099 during the flare and (d) HR 1099 during the quiescence. Blue and red Gaussians in panel a correspond to Fe K$\alpha_{1}$ and K$\alpha_{2}$ components, respectively. The green Gaussian in panel c corresponds to the sum of Fe K$\alpha_{1}$ and K$\alpha_{2}$ lines. The spectra are binned only for display. 
{Alt text: Four line graphs show the spectra and residuals to the best-fit models around the Fe K-alpha line for GT Mus, Sigma Gem, and HR 1099.}
}
\label{fig:FeKa}
 \end{center}
\end{figure*}

Here, we update previous Fe~K$\alpha$ modeling efforts described above for the microcalorimeter era by using the three-dimensional Monte Carlo radiative-transfer code \texttt{SKIRT} v9.0 \citep{Vander_2023}.
Earlier studies have often adopted geometrical setup tailored for accreting systems.
In this paper, we take late-type stars instead. 
A key geometric advantage of this system is that it is simple as there is no thick target such as an accretion disk or stellar wind.
The photosphere is the only reflector, so the geometry is simple and can be parameterized primarily by $h/R_{*}$, where $R_{*}$ is the stellar radius and $h$ is the height of the incident point source above the stellar surface.

In fact, the Fe~K$\alpha$ line is frequently detected in solar and stellar X-ray spectra, particularly during flares \citep{Neupert_1967, Neupert_1971, Doschek_1971, Bai_1979, Feldman_1980, Culhane_1981, Parmar_1982, Tanaka_1982, Tanaka_1983, Tanaka_1984, Parmar_1984, Tanaka_1985, Emslie_1986, Zarro_1992, Phillips_1995, Osten_2007, Testa_2008, Osten_2010, Huenemoerder_2010, Karmakar_2017, Inoue_2025, Suzuki_2025, Inoue_2026, Sarwade_2026}.
Its dominant production mechanism is generally attributed to photoionization of the lower atmosphere (transition region, chromosphere, and photosphere) by hard X-rays from flare loops \citep{Bai_1979, Inoue_2025, Inoue_2026, Suzuki_2025}.
By representing this scenario as an X-ray point source (a flare loop or a coronal bright point) illuminating a spherical reflector (the photosphere) and performing three-dimensional radiative-transfer calculations, we present Fe~K$\alpha$ equivalent-width maps that are readily applicable to other systems.

The plan of this paper is as follows. We first describe the XRISM/Resolve observations of three late-type stars (Section~\ref{sec:data}), and then present the setup and results of our \texttt{SKIRT} radiative-transfer calculations (Section~\ref{sec:RT}).
In Section~\ref{sec:discussion}, we discuss the results by comparing the observations with the calculations, and we summarize our conclusions in Section~\ref{sec:conclusion}.
Unless otherwise stated, all quoted uncertainties correspond to the 90\% confidence level.

\section{Data}\label{sec:data}
Among late-type stars, RS~Canum Venaticorum (RS CVn) type stars are the most magnetically active and X-ray luminous sources \citep[e.g.,][]{Oord_1988, Tsuru_1989, Doyle_1991, White_1994, Osten_2000, Franciosini_2001, Osten_2003, Osten_2007, Pandey_2012, Tsuboi_2016, Sasaki_2021, Kurihara_2024, Inoue_2024b, Inoue_2025, Inoue_2026, Urabe_2026}, providing high-statistics X-ray spectra and making them ideal targets for this study.
As of January 2026, XRISM has obtained observations of three RS~CVn-type stars---GT~Muscae (GT~Mus), $\sigma$~Geminorum ($\sigma$~Gem), and HR~1099---and clear evidence for a flare has been found only in the HR~1099 observation.
Table~\ref{tab:best-fit} summarizes the data set used in this study.

\subsection{Target}\label{subsec:Target}

\subsubsection{GT Mus}\label{subsubsec:GT_Mus}
GT~Mus is a quadruple system \citep{Collier_1982} consisting of HD~101379 and HD~101380, located at a distance of 141~pc \citep{Gaia_2016, Gaia_2023}.
HD~101379 is an RS~CVn-type single-lined spectroscopic binary \citep{Strassmeier_1988, McAlister_1990}, consisting of the G5--8 giant and an A0--1 dwarf \citep{Houk_1975}.
The G5--8 giant primary of HD~101379 (GT~Mus~Aa) has a radius of $R_{*}=16.6\,R_{\odot}$ \citep{Kallinger_2019}, exhibits strong coronal activity, and has produced large stellar flares \citep{Jones_1995, Tsuboi_2016, Sasaki_2021, Eze_2022}.
HD~101380 is an eclipsing binary composed of A0 and A2 dwarfs \citep{Murdoch_1995} and is expected to contribute little to the total X-ray emission of the system.

\subsubsection{$\sigma$ Gem}\label{subsubsec:Sigma_Gem}
$\sigma$~Gem is a single-lined spectroscopic binary \citep{Herbig_1955, Huenemoerder_2013, Roettenbacher_2015b} consisting of a K1 giant primary \citep{Roman_1952} and a G or K dwarf secondary \citep{Ayres_1984, Kovari_2001}, located at a distance of 38.3~pc \citep{Roettenbacher_2015}.
The K-type primary has a mass of $M=1.28\,M_{\odot}$ and a radius of $R=10.1\,R_{\odot}$ \citep{Roettenbacher_2015b}.
$\sigma$~Gem has been observed in X-rays on multiple occasions \citep[e.g.,][]{Singh_1987}, and flares have been reported across a wide range of wavelengths \citep[e.g.,][]{Brown_2006, Nordon_2006, Pandey_2012}.

\subsubsection{HR 1099}\label{subsubsec:HR1099}
HR~1099 is a hierarchical binary consisting of a K3 dwarf and an RS CVn-type binary of a K1 subgiant and a G5 giant \citep{Boop_1976}.
The primary (HR~1099~Aa) has a radius of $R=3.7\,R_{\odot}$ and a mass of $M=1.0\,M_{\odot}$ \citep{Donati_1999}.
With its strong coronal activity and proximity \citep[29.4 pc;][]{Gaia_2021}, HR~1099 is one of the best-known RS~CVn-type stars and has served as a calibration target for many X-ray telescopes, including XRISM \citep[e.g.,][]{Marshall_2004, Eckart_2025}.
Moreover, owing to its high coronal activity and frequent observations, numerous flares from HR 1099 have been reported across a broad range of wavelengths \citep[e.g.,][]{Trigilio_1993, Umana_1995, Osten_1999, Ayres_2001, Audard_2001, Osten_2004, Nordon_2007, Didel_2025, Wang_2026}.

\subsection{Observation and Data Reduction}\label{subsec:Observation_Reduction}

GT~Mus was observed with XRISM as a performance-verification target (sequence number 300000010) from 2024 August 13 02:03:40 to August 17 10:12:28~UT, when the source was in quiescence, with a total exposure of 199~ks (Figure~\ref{fig:GTMus_lc}a).
Details of the observation and data set are provided by \citet{Kurihara_2025}.

$\sigma$~Gem was observed with XRISM as a Cycle~2 Guest Observer (GO) target (sequence number 202063010) from 2025 November 06 22:19:04 to November 09 08:52:04~UT, when the target was in quiescence, with a total exposure of 109~ks (Figure~\ref{fig:GTMus_lc}b).
These data were obtained as part of the multi-wavelength campaign, which also included simultaneous optical spectroscopy with the Nayuta telescope at the Nishi-Harima Astronomical Observatory and radio observations with the Yamaguchi Radio Telescope \citep{Fujisawa_2002}.
We only focus on the analysis of the Fe K$\alpha$ line in this paper. 
Our results using other X-ray features and optical and radio wavelength data will be presented in upcoming papers.

HR~1099 was observed with XRISM from 2024 March 06 01:21:11 to March 10 22:40:02~UT as an instrument-calibration target.
Details of the observation and data set are provided by \citet{Kurihara_2025b}.
Figure~\ref{fig:GTMus_lc}c shows the Resolve light curve of HR~1099.
A clear flare was detected around $t\simeq 2$~day, with the count rate increasing by a factor of $\sim 8$ and a total radiated energy of $\sim 10^{34}$~erg \citep{Kurihara_2025b}.
To date, this is the only flare from an RS~CVn-type star observed by Resolve.

The data of GT Mus and $\sigma$ Gem were processed and calibrated using the standard pipeline with HEASoft v6.35.1 \citep{heasoft_2014} and the calibration database (\texttt{CALDB}) version \texttt{20250315}.
The HR~1099 data obtained in the calibration observation underwent several special reduction procedures dedicated for instrumental setup for calibration purposes, the details of which are summarized in \citet{Kurihara_2025b} and \citet{Eckart_2025}.
For all three sources, we extracted spectra from 33 pixels, excluding pixel 12 used for calibration and pixel 27, which has occasional anomalies in gain variation characteristics, and used high-primary (Hp) grade events only.
We did not subtract a background spectrum because it is negligible in the energy band of interest \citep{Kilbourne_2018, Mochizuki_2025}.
For HR 1099, we extracted spectra separately for the flare interval and the remaining time, following \citet{Kurihara_2025b}.

\subsection{Analysis}\label{subsec:analysis}
\subsubsection{Narrow-band}\label{subsubsec:Narrow-band}
Figure~\ref{fig:FeKa} shows the narrow-band spectra (6.30--6.47~keV) around the Fe~K$\alpha$ line.
First, we fitted the narrow-band spectrum with the phenomenological two Gaussian components to represent the Fe~K$\alpha_{1}$ ($2p_{3/2} \rightarrow 1s$) and K$\alpha_{2}$ ($2p_{1/2} \rightarrow 1s$) lines plus the power-law component to account for the underlying continuum using the Cash statistic \citep[][]{Cash_1979}.
We set the fitting parameters according to the following procedures:
\begin{enumerate}
    \item The photon index of the power-law component was fixed at 0, while the normalization was left free.
    \item The Gaussian line widths were initially allowed to vary independently. If they could not be constrained, we tied the widths of the two components to be equal. If the width still remained unconstrained, we fixed it at zero.
    \item If the 90\% lower bound on the normalization of either Gaussian component was below zero, that component was removed and the spectrum was refitted with a single Gaussian line.
    \item If the 90\% lower bound on the normalization of the single Gaussian was still below zero, we fixed the centroid energy to 6.4~keV and the line width to zero.
\end{enumerate}
After obtaining the best-fit model following these procedures, we computed the Fe~K$\alpha$ equivalent width or the 90\% upper limit on it.
The uncertainties and 90\% upper limits of the equivalent widths were estimated with the \texttt{error} command in \texttt{Xspec} by searching for the parameter range corresponding to the 90\% confidence interval based on the increase in the C-statistic from the best-fit value.

The GT~Mus spectrum shows a clear detection of the two Fe~K$\alpha$ emission lines (Figure \ref{fig:FeKa}a).
The best-fit centroid energies are $6.40^{+0.00}_{-0.01}$~keV and $6.39^{+0.01}_{-0.02}$~keV, consistent with the laboratory energies of Fe~K$\alpha_{1}$ and K$\alpha_{2}$ lines, respectively \citep{Bearden_1967, Nagai_2026}.
The common line width is $3.9^{+2.3}_{-1.2}$~eV, corresponding to a velocity broadening of $180^{+110}_{-60}$~km~s$^{-1}$.
The line-flux ratio is $F_{\mathrm{K\alpha_{1}}}/F_{\mathrm{K\alpha_{2}}}=2.2^{+1.0}_{-1.8}$, consistent with the expected branching ratio of $\mathrm{K\alpha_{1}:K\alpha_{2}}=2{:}1$.
The equivalent widths of Fe~K$\alpha_{1}$ and K$\alpha_{2}$ lines are $16.2^{+5.0}_{-6.6}$~eV and $7.5^{+3.8}_{-5.5}$~eV, respectively, and the combined equivalent width is $\mathrm{EW}_{\mathrm{K\alpha}} = 23.8^{+5.7}_{-8.3}$~eV.
To date, the only reported detection of stellar Fe~K$\alpha$ emission without an obvious flare is the NICER detection in November 2017 \citep{Inoue_2025}, also for GT~Mus. However, the equivalent width reported by \citet{Inoue_2025} is $160 \pm 107.6$~eV, nearly five times larger than the value measured here.

No Fe~K$\alpha$ line was detected in the $\sigma$~Gem spectrum (Figure \ref{fig:FeKa}b).
The derived 90 \% upper limit on the Fe~K$\alpha$ equivalent width is $\mathrm{EW}_{\mathrm{K\alpha}}<3.9$~eV.

For the HR 1099 spectrum during the flare, the photon statistics were insufficient to separate the K$\alpha_1$ and K$\alpha_2$ components. 
We detected Fe~K$\alpha$ with a centroid energy of $E_{l}=6.40^{+0.01}_{-0.02}$~keV and an equivalent width of $\mathrm{EW}_{\mathrm{K\alpha}}=4.6^{+3.7}_{-2.9}$~eV (Figure~\ref{fig:FeKa}c).
In contrast, no Fe~K$\alpha$ line was detected in the quiescent spectrum (Figure~\ref{fig:FeKa}d).
We therefore derived 90 \% upper limit of $\mathrm{EW}_{\mathrm{K\alpha}}<3.6$~eV.

\begin{table}[]
\centering
\caption{Best-fit temperatures and 1.7--10~keV fluxes.}
\label{tab:bestfit_kT_flux}
\renewcommand{\arraystretch}{1.25}
\begin{tabular}{cccc}
\hline
Star & Phase & $kT$ & $F_{\mathrm{X}}$ \\
& & (keV) & ($10^{-11}$~erg~cm$^{-2}$~s$^{-1}$) \\
\hline
GT~Mus & Quiescence & $2.64^{+0.04}_{-0.03}$ & $2.62^{+0.02}_{-0.02}$  \\ \hline
$\sigma$~Gem & Quiescence & $2.07^{+0.03}_{-0.03}$ & $2.75^{+0.03}_{-0.02}$  \\ \hline
\multirow{2}{*}{HR~1099}
& Flare & $2.65^{+0.05}_{-0.05}$ & $5.31^{+0.05}_{-0.06}$ \\
& Quiescence & $1.94^{+0.03}_{-0.03}$ & $3.08^{+0.02}_{-0.03}$ \\
\hline
\end{tabular}
\end{table}

\subsubsection{Broad-band}\label{subsubsec:Broad-band}
As the Fe K$\alpha$ line is the response to the incident X-ray photons above the Fe K edge energy, it is important to characterize the continuum shape.
We therefore fitted the broad-band (1.7--10.0~keV) spectra with a single-temperature thermal plasma model without emission lines \citep{Smith_2001}, leaving both the temperature and the normalization free.
The best-fit parameters are summarized in Table~\ref{tab:bestfit_kT_flux}.


\begin{figure}[]
\begin{center}
\includegraphics[width=0.46\textwidth]{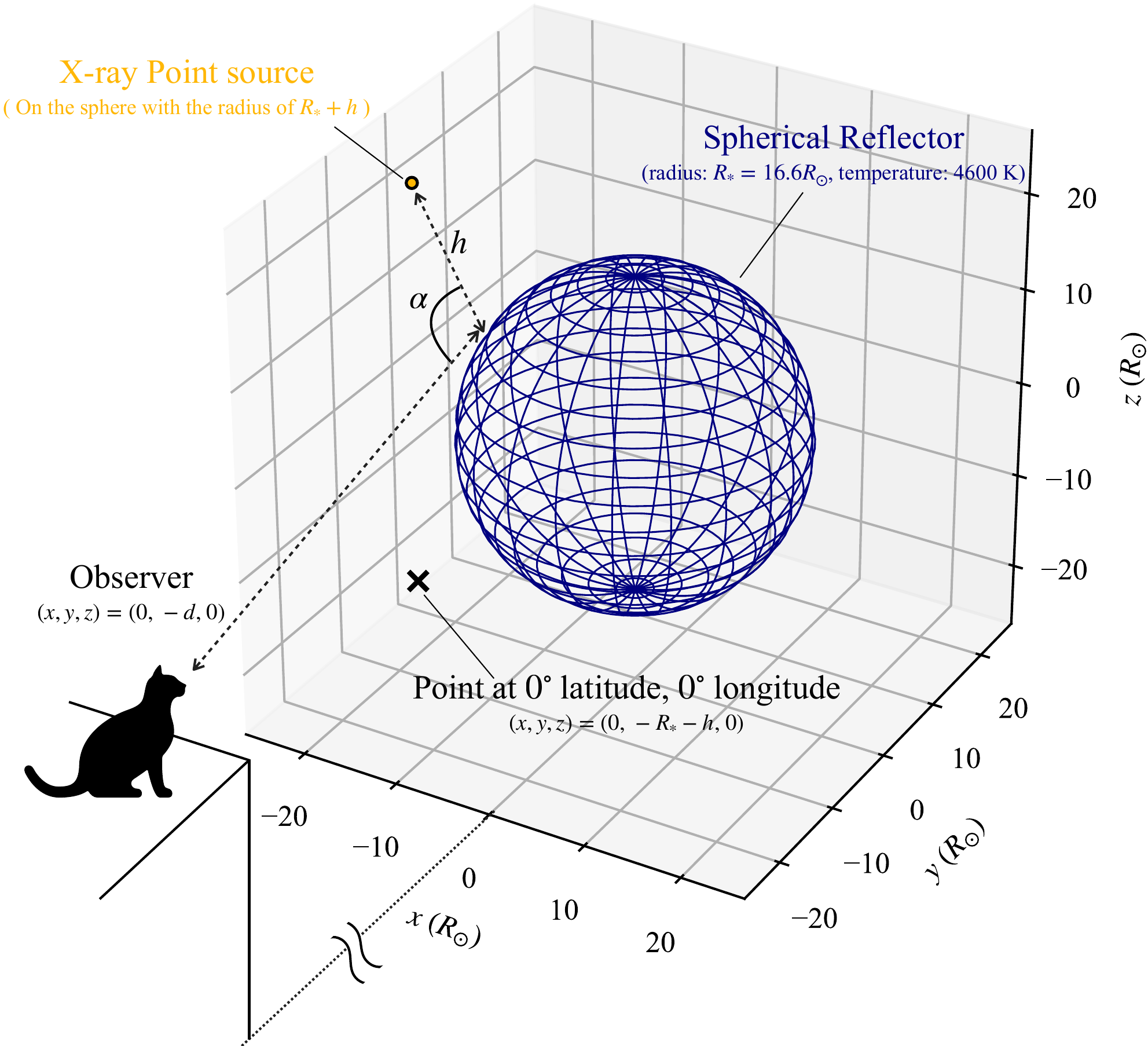}
\caption{Schematic diagram of the setup of the \texttt{SKIRT} simulation.
{Alt text: Schematic diagram showing the geometric setup used in the SKIRT simulation.}
}
\label{fig:calc_setup}
 \end{center}
\end{figure}

\begin{figure*}[]
\begin{center}
\includegraphics[width=0.95\textwidth]{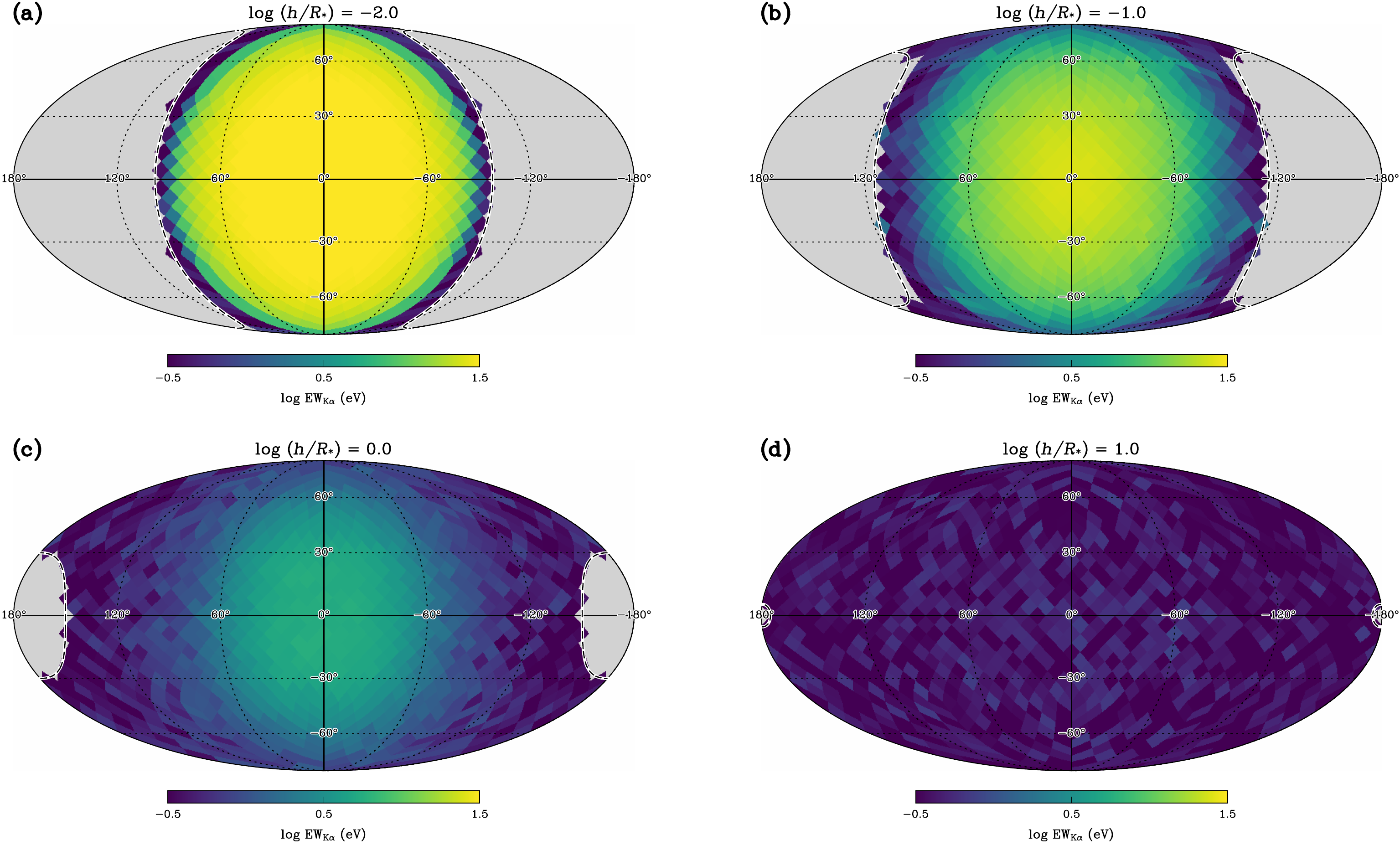}
\caption{Fe K$\alpha$ equivalent width maps obtained by \texttt{SKIRT} for $\log (h/{R_{*}}) = -2$, $-1$, $0$, and $1$ cases. The black dashed line shows the limb.   
Note that these are results when the GT Mus quiescence spectrum is used as an input SED.
{Alt text: Four color maps show the Fe K-alpha equivalent-width maps for different source heights, with the stellar limb indicated by a dashed line.}
}
\label{fig:Mollweide_map}
 \end{center}
\end{figure*}

\begin{figure}[]
\begin{center}
\includegraphics[width=0.44\textwidth]{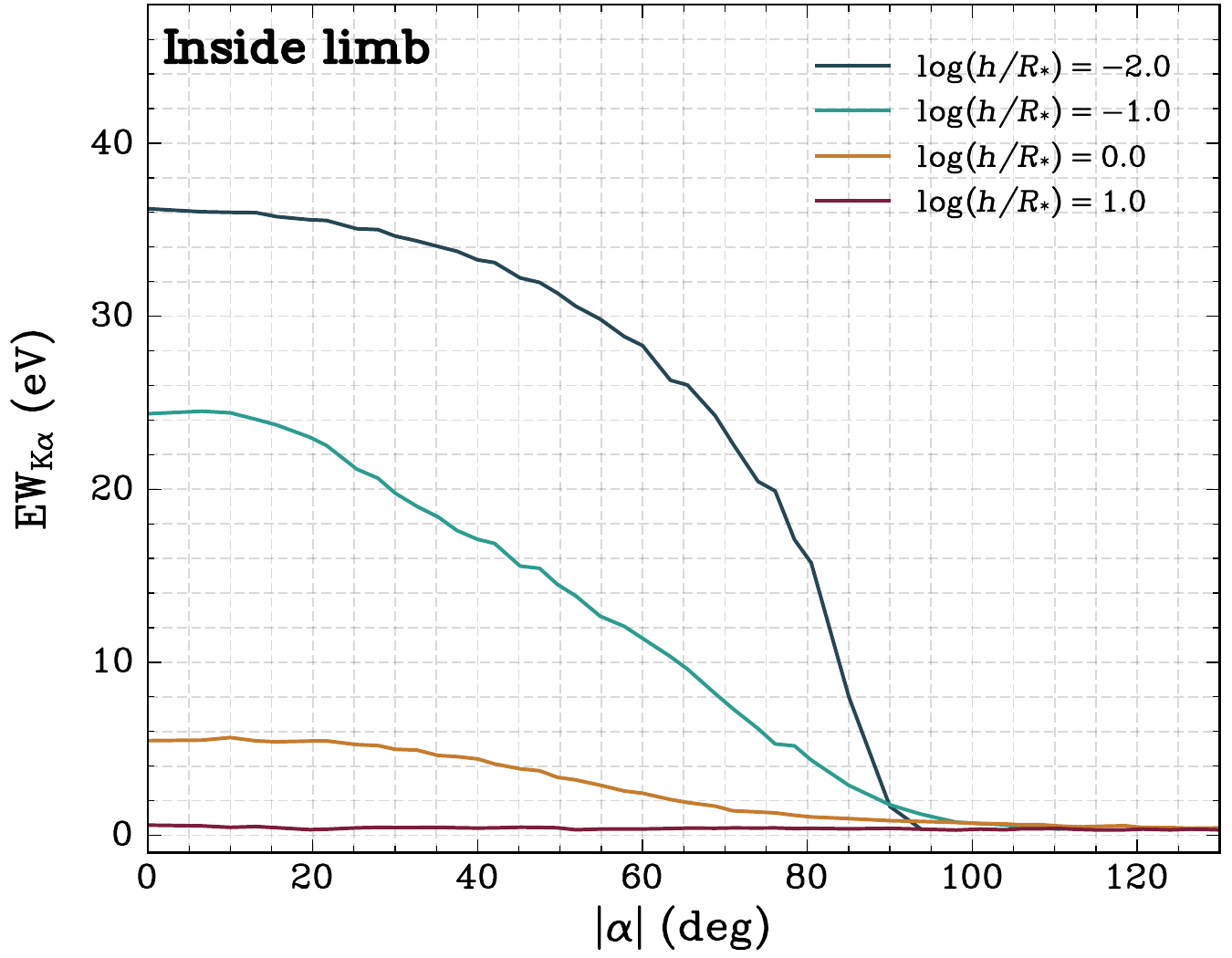}
\caption{
Fe K$\alpha$ equivalent width as a function of $|\alpha|$. 
Blue, green, orange, and red lines correspond to $\log(h/R_{*}) = -2, -1, 0$ and 1, respectively.
{Alt text: One line graph for the Fe K-alpha equivalent width as a function of the angular separation.}
}
\label{fig:inclination_vs_ew}
 \end{center}
\end{figure}

\section{Radiative Transfer Calculation}\label{sec:RT}
\texttt{SKIRT} is an open-source Monte Carlo radiative-transfer (MCRT) code originally developed for galaxy studies \citep{Baes_2003, Baes_2011, Camps_2015, Baes_2015, Camps_2020}, but it has also been applied to a variety of other systems thanks to its rich set of built-in options for geometries, radiation sources, transfer-medium properties, spatial grids, and detectors \citep[e.g.,][]{Toala_2024b, Mochizuki_2024, Toala_2025, Inoue_2026, Fujiwara_2026, Fujiwara_2026b, Hamaguchi_2026}.
A key advantage of \texttt{SKIRT} over other MCRT codes is that it fully incorporates importance-sampling techniques \citep{Baes_2016}, such as photon peel-off \citep[e.g.,][]{Yusef-Zadeh_1984}, enabling calculations with a complex geometrical setup and high-energy resolution to be completed within practical run times.
X-ray radiative processes implemented in \texttt{SKIRT} v9.0 include electron scattering, either free or bound, photoelectric absorptionn, fluorescence by the neutral matter with its intrinsic line profiles, dust extinction, and dust scattering \citep{Vander_2023, Vander_2024}.

We performed three-dimensional radiative-transfer calculations with \texttt{SKIRT} v9.0 \citep{Vander_2023}.
Our goal is not only to infer the geometry of coronal X-ray sources from Fe~K$\alpha$ observations of RS~CVn-type stars (Section~\ref{sec:data}), but also to construct Fe~K$\alpha$ equivalent-width maps in a sufficiently generalized framework so that they can be applied to a wide variety of other systems.

\subsection{Setup}\label{subsec:setup}
Figure~\ref{fig:calc_setup} shows a schematic of our simulation setup.
We consider a generalized system consisting of a spherical reflector of radius $R_{*}$ and an X-ray point source located at a height $h$ above the surface.
Longitude $0^\circ$ is defined by the great-circle arc passing through $(0,0,-R_{*}-h)$, $(0,-R_{*}-h,0)$, and $(0,0,R_{*}+h)$, corresponding to the direction directly facing the observer, whereas longitude $180^\circ$ is defined by the great-circle arc passing through $(0,0,-R_{*}-h)$, $(0,R_{*}+h,0)$, and $(0,0,R_{*}+h)$.
We define $\alpha$ as the angular separation on the spherical reflector between the observer and the location of the X-ray point source. 
The only essential parameter for the geometry is $h/R_{*}$.

For the actual values, we use GT~Mus as an example because it shows the largest Fe~K$\alpha$ equivalent width among our targets (Table~\ref{tab:best-fit}); we therefore set $R_{*}=16.6\,R_{\odot}$, the radius of GT~Mus~Aa \citep{Kallinger_2019}.
We adopt solar photospheric values for the elemental abundances and physical conditions of the sphere: the abundance pattern is set to $Z_{\odot}$ \citep{Anders_1989}, the proton number density to $n_{\mathrm{p}}=10^{16}~\mathrm{cm^{-3}}$ \citep{Bommier_2020}, and the temperature to the effective temperature of GT~Mus, $T_{\mathrm{eff}}=4740$~K \citep{Gaia_2018}.
The best-fit continuum model of GT~Mus (Section~\ref{subsubsec:Broad-band}) was adopted as the input source SED.
The observer's position and line of sight were fixed at $(0,-d,0)$ and along the $+y$ direction, respectively, where $d=141$~pc \citep{Gaia_2016, Gaia_2023} is the distance from GT~Mus to Earth.
The number of seed pseudo-photons was set to $10^{8}$.

Under these assumptions, we performed radiative-transfer calculations for four heights, $\log(h/R_{*})=-2$, $-1$, $0$, and $1$.
In order to make a grid of an equal area over the sphere, we utilized \texttt{HEALPix} \citep{Gorski_2005} and its Python module \texttt{healpy} \citep{Zonca_2019} commonly used to describe the anisotropy of the cosmic microwave background \citep[e.g.,][]{Bortolami_2025}.
We tessellated the sphere of radius $R_{*}+h$ into $N_{\mathrm{pix}}=12\,N_{\mathrm{side}}^{2}=972$ equal-area pixels ($N_{\mathrm{side}}=9$), and computed the observed spectrum for a source placed at the center of each pixel.
We computed the Fe~K$\alpha$ equivalent width after convolving the \texttt{SKIRT}-simulated spectra to the Resolve energy resolution, $R=1400$ at 6.4~keV.

In this paper, we define ``limb'' on the sphere of radius $R_{*}+h$ (black dashed curves in Figure \ref{fig:Mollweide_map}).
Outside the limb, the source is occulted by the sphere and is not directly visible to the observer. Inside the limb, the source is directly visible.
We also define the ``disk'' as the portion of the photosphere visible to the observer, i.e., the projected stellar surface within the limb.
In this calculation, we consider only cases in which the source is placed inside the limb. If the source is located outside the limb, some photons can reach the observer after skimming through the stellar surface layers. A rigorous treatment of such configurations would require explicitly modeling the density structure of the atmospheric surface, which would reduce the generality of the results. 
Therefore, we leave such a specific example in Section \ref{subsec:realistic} and proceed here without considering the situation in the limb.

The computation was performed on the supercomputer of ACCMS, Kyoto University, using 896 CPU cores ($=$ 112 CPU cores $\times$ 8 nodes). 
We used the hybrid parallelization capability of SKIRT, in which multiple execution threads are assigned to each process\footnote{\url{https://skirt.ugent.be/root/_user_parallel.html}}. 
Specifically, we assigned 16 threads to each SKIRT process and ran 7 processes on each 112-core node. We performed the calculations on 8 nodes in parallel.
As $h$ decreases, individual photons penetrate deeper into the sphere and undergo more interactions, increasing the computational cost.
For the most computationally demanding case, $\log(h/R_{*})=-2$, the production run required $38$ hours of wall-clock time, corresponding to a total of $34000$ core-hours of CPU time.


\begin{figure}[t]
\begin{center}
\includegraphics[width=0.45\textwidth]{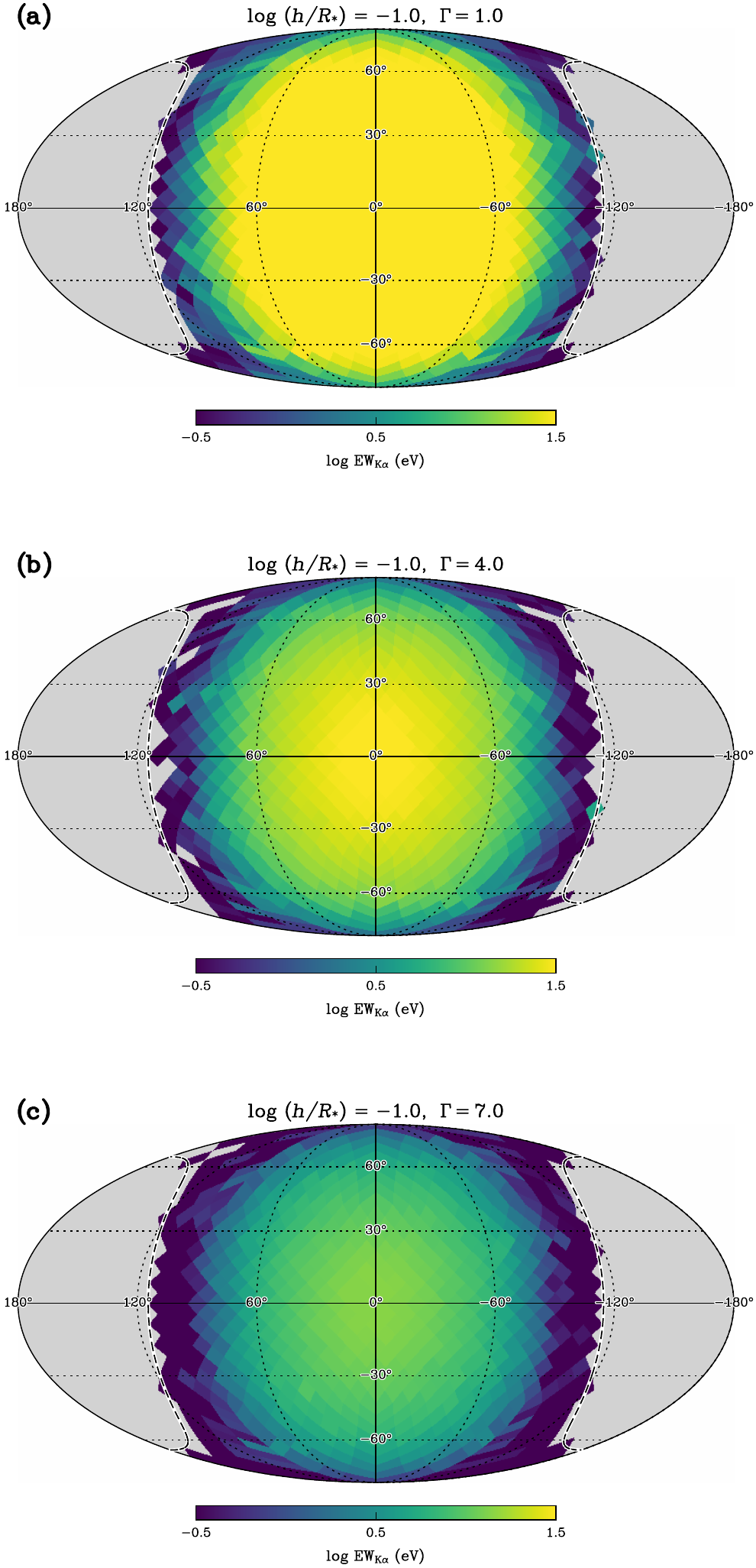}
\caption{Equivalent-width maps for $\log (h/R_{*}) = -1$ when powerlaw spectra with the indices of (a) 1.0 ,(b) 4.0, and (c) 7.0 are used as an input SED.
{Alt text: Three color maps show the Fe K-alpha equivalent-width maps for different powerlaw indices.}
}
\label{fig:Mollwide_powerlaw}
 \end{center}
\end{figure}

\subsection{Results}\label{sec:RT_results}
Figure~\ref{fig:Mollweide_map} shows Fe~K$\alpha$ equivalent-width maps from our \texttt{SKIRT} calculations, plotted in a Mollweide projection on the sphere of radius $R_{*}+h$.
The color scale of each pixel represents the Fe~K$\alpha$ equivalent width for a source placed at the pixel center.
The black dashed line indicates the limb defined in Section \ref{subsec:setup}.

The Fe~K$\alpha$ equivalent-width maps (Figure~\ref{fig:Mollweide_map}) exhibit two characteristic features:
\begin{enumerate}
    \item an increase in equivalent width toward the disk center;
    \item an overall decrease in equivalent width with increasing $h/R_{*}$.
\end{enumerate}

First, the feature toward the disk center reflects the increase in the solid angle, as seen by the observer, subtended by the Fe~K$\alpha$-emitting region as the source approaches the disk center.
This trend has indeed been observed for both Fe~K$\alpha$ and K$\beta$ lines in solar flares \citep{Parmar_1982, Parmar_1984, Phillips_1994, Phillips_1995, Phillips_2012}.

Second, the overall decrease in equivalent width with increasing $h/R_{*}$ reflects the decline of the fluorescence efficiency with increasing the loop height $h$ due to the $1/h^{2}$ geometrical dilution of the hard X-ray flux at stellar surface.
This behavior is consistent with previous theoretical and observational studies of solar and stellar Fe~K$\alpha$ emission \citep{Bai_1979, Drake_2008, Ercolano_2008, Testa_2008, Inoue_2025, Inoue_2026, Sarwade_2026}.

The Fe~K$\alpha$ equivalent-width maps are symmetric about the point at 0$^{\circ}$ latitude and 0$^{\circ}$ longitude, and can therefore be parameterized solely by $|\alpha|$. 
Figure \ref{fig:inclination_vs_ew} shows the Fe~K$\alpha$ equivalent width as a function of $|\alpha|$, obtained by taking the median value over all pixels at each $|\alpha|$.
The equivalent-width range shown in Figure \ref{fig:inclination_vs_ew} ($<40$ eV) corresponds to the regime that has become accessible thanks to the improved detection threshold from $\sim 50$~eV with Chandra/HETG to $\sim 5$~eV with XRISM/Resolve.

\begin{table}[]
\centering
\caption{The median equivalent width for $\log (h/R_{*}) = -1$ inside the limb when the GT Mus quiescence spectrum and powerlaw spectra with the indices of 1.0, 4.0, and 7.0 are used.} 
\label{tab:med}
\renewcommand{\arraystretch}{1.2}
\begin{tabular}{ccc}
\hline
\multicolumn{2}{c}{input SED} & Med (EW$_{\mathrm{K\alpha}}$)   \\ 
 & & (eV) \\ \hline
Bremsstrahlung            & $kT = 2.64$ keV & 7 \\ \hline
\multirow{3}{*}{powerlaw} & $\Gamma = 1.0$  & 31 \\
                          & $\Gamma = 4.0$  & 10 \\
                          & $\Gamma = 7.0$  & 4 \\ \hline
\end{tabular}
\end{table}

\section{Discussion}\label{sec:discussion}
\subsection{Dependence of the maps on the input SED}
The maps shown in Figure~\ref{fig:Mollweide_map} are geometrically generalized by normalizing $h$ to $R_{*}$. They are also generalized with respect to the source X-ray luminosity by presenting equivalent-width maps rather than line-luminosity maps.
The only aspect that is not fully general is the input SED.

The Fe~K$\alpha$ equivalent-width maps in Figure~\ref{fig:Mollweide_map} were computed using the quiescent GT~Mus spectrum as the input SED. This spectrum is well approximated by thermal bremsstrahlung from a plasma with $kT=2.64$~keV (Section \ref{subsubsec:Broad-band}).
The thermal-bremsstrahlung photon spectrum can be written as
\begin{equation}
    \frac{dN}{dE} \propto E^{-1}\exp(-E/kT).
\end{equation}
The effective photon index obtained by locally approximating the spectrum as a power law, $dN/dE \propto E^{-\Gamma_{\mathrm{eff}}}$, is
\begin{equation}
\label{eq:Gamma_eff}
    \Gamma_{\mathrm{eff}} \equiv - \frac{d \log(dN/dE)}{d \log E} = 1 + E/kT .
\end{equation}
Substituting $kT=2.64$~keV and $E=6.4$~keV into Equation~\ref{eq:Gamma_eff} yields $\Gamma_{\mathrm{eff}}=3.4$, which characterizes the input SED at the Fe~K$\alpha$ energy for the maps in Figure~\ref{fig:Mollweide_map}.
To assess the sensitivity of the maps to the input SED, we repeated the $\log(h/R_{*})=-1$ calculations using power-law spectra with photon indices of $\Gamma=1.0$, 4.0, and 7.0.

Figure~\ref{fig:Mollwide_powerlaw} shows how the equivalent-width maps change when the power-law index is varied.
The absolute equivalent width changes with the power-law index because the ratio between the photon flux above the Fe K edge and the continuum flux at 6.4~keV depends on the spectral slope.
In contrast, the spatial distribution depends only on $h/R_{*}$, so varying the index mainly appears as an overall shift in the equivalent-width normalization.
Table~\ref{tab:med} summarizes the median equivalent width.
The median equivalent width is approximately inversely proportional to the power-law index, consistent with the results of \citet{George_1991}.
Therefore, when applying the maps shown in Figure~\ref{fig:Mollweide_map} to other objects, one should keep in mind an overall uncertainty of a factor of a few due to differences in the input SED.

\begin{figure}[]
\begin{center}
\includegraphics[width=0.45\textwidth]{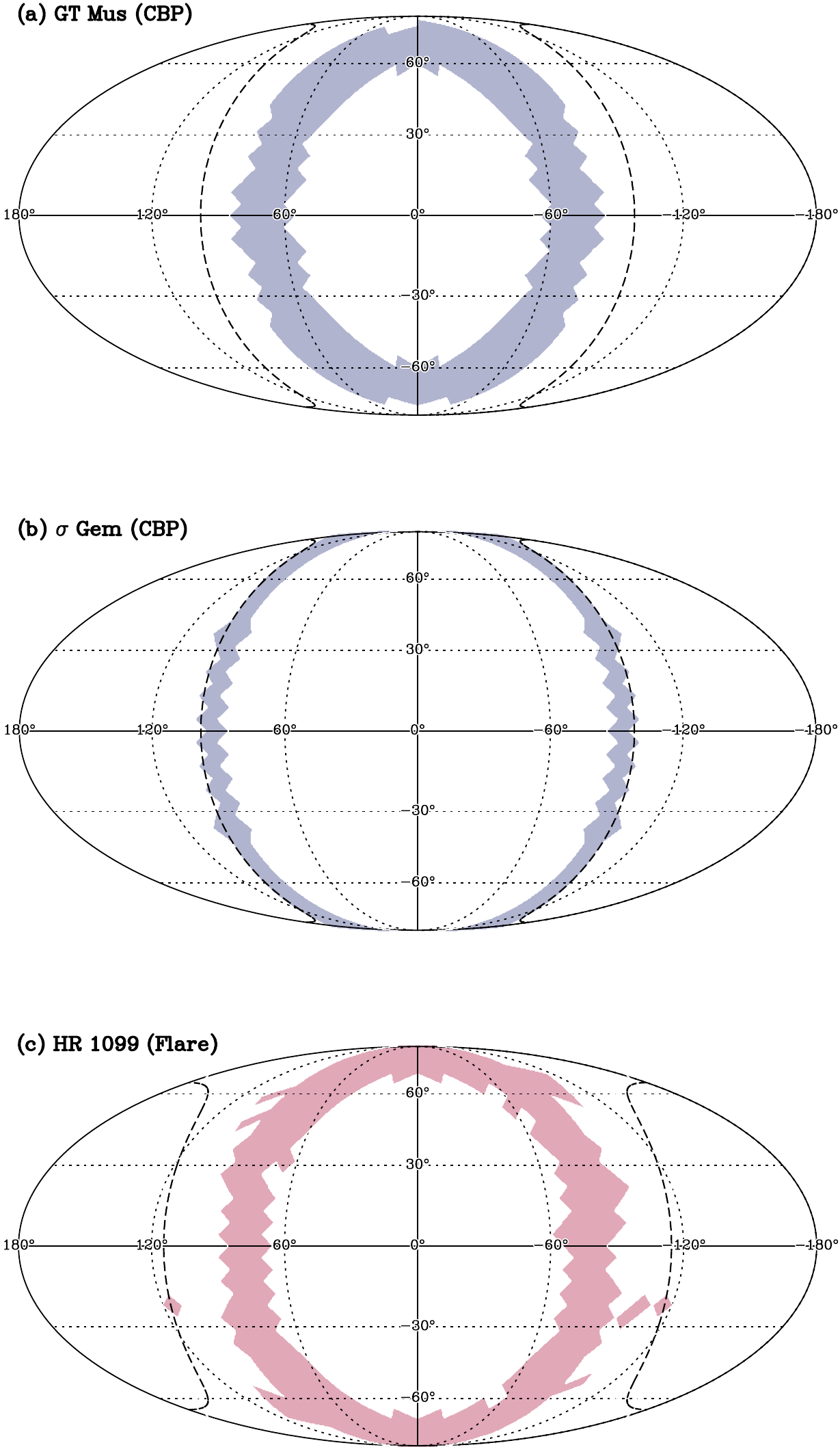}
\caption{Location of coronal X-ray sources of (a) GT Mus, (b) $\sigma$ Gem, and (c) HR 1099 on the sphere with the radius of $R_{*}+h$. Each region corresponds to the 90 \% error range of the source location. Note that $\log (h/R_{*}) = -2$ for panel a and b and $\log (h/R_{*}) = -1$ for panel c.
{Alt text: Three maps show the inferred coronal X-ray source locations for GT Mus, Sigma Gem, and HR 1099.}
}
\label{fig:Mollwide_location}
 \end{center}
\end{figure}

\begin{figure}[]
\begin{center}
\includegraphics[width=0.45\textwidth]{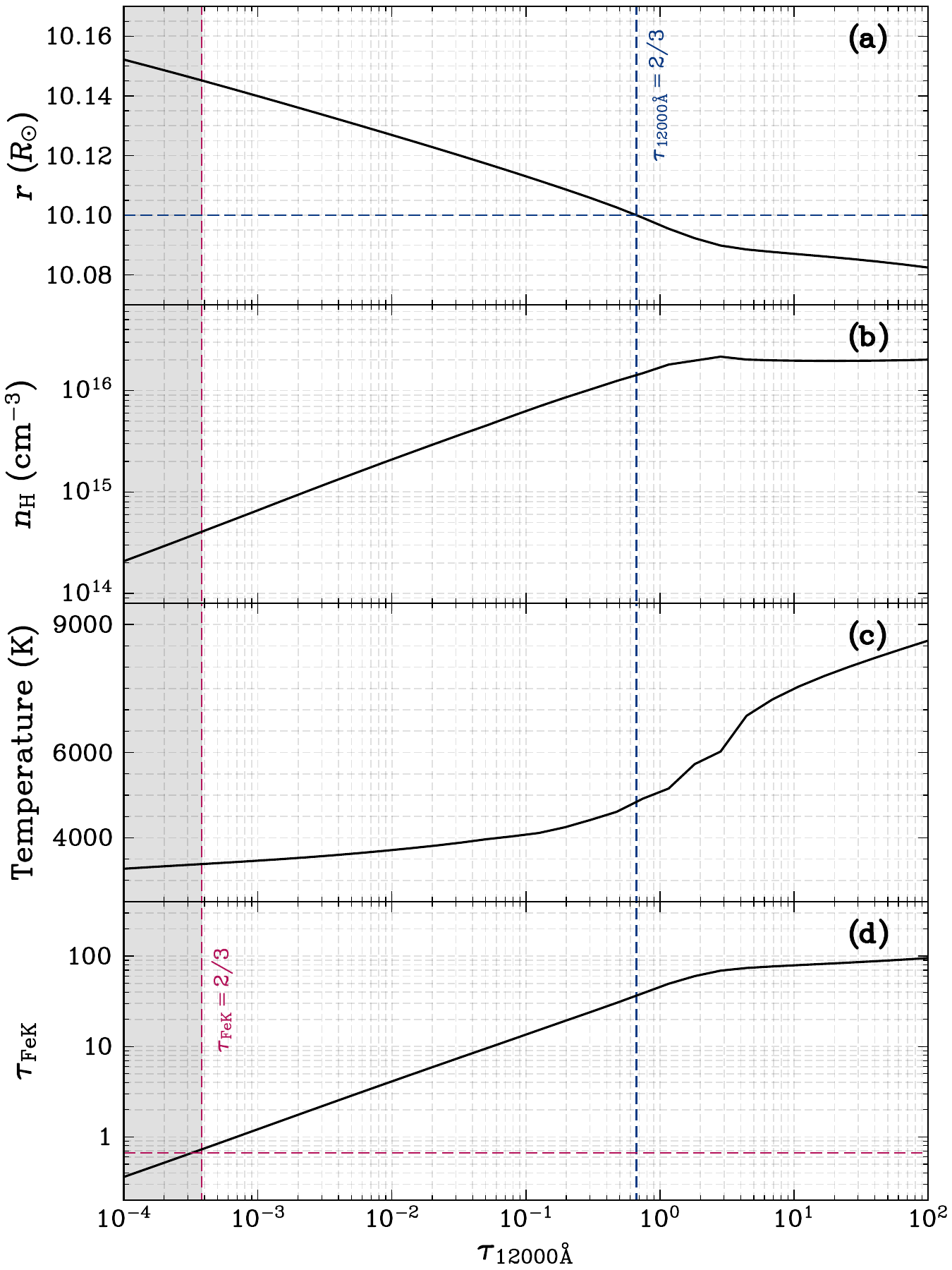}
\caption{Atmosphere model for the effective temperature of 4600 K and the surface gravity of $10^{2.5}$ cm s$^{-2}$ based on the calculations of \texttt{PHOENIX} \citep{Husser_2013}. (a) the distance from the center of the star, (b) hydrogen density, (c) temperature, and (d) the Fe K optical depth ($\tau_{\mathrm{FeK}}$) as a function of optical depth at 12000 $\mathrm{\mathring{A}}$ ($\tau_{\mathrm{12000 \mathring{A}}}$). Blue and red dashed lines correspond to $\tau_{\mathrm{12000 \mathring{A}}}=2/3$ and $\tau_{\mathrm{FeK}} = 2/3$, respectively. The data available online at <\url{https://phoenix.astro.physik.uni-goettingen.de}>.
{Alt text: Four line graphs of the stellar atmosphere model for the distance from the center of the star, hydrogen density, temperature, and Fe K optical depth as a function of optical depth at 12000 angstrom.}
}
\label{fig:phoenix}
 \end{center}
\end{figure}

\subsection{Location of coronal X-ray sources}\label{subsec:location}
We can constrain the locations of the coronal X-ray sources by comparing the Fe~K$\alpha$ measurements for the RS~CVn-type stars (Figure~\ref{fig:FeKa}) with the equivalent-width maps (Figure~\ref{fig:Mollweide_map}), thereby illustrating how the maps can be used in practice.

For GT~Mus and $\sigma$~Gem, in which no clear flare was detected during our observations (Figure~\ref{fig:GTMus_lc}), we assume that the coronal X-ray source corresponds to a localized active region, i.e., a coronal bright point (CBP), as observed on the Sun \citep[e.g.,][]{Nitta_1988, Nitta_1992, Strong_1992, Hara_1992, Harvey_1993, Shimojo_1999, Hara_2003, Kotoku_2007, Tian_2008, Hara_2009, Madjarska_2019, Paterson_2024, Paterson_2026}.
A typical loop height of solar CBPs is $h \sim 0.1\,R_{\odot}$ \citep[e.g.,][]{Madjarska_2024}, while the stellar radii are $R_{*}=16.6\,R_{\odot}$ for GT~Mus \citep{Kallinger_2019} and $R_{*}=10.1\,R_{\odot}$ for $\sigma$~Gem \citep{Roettenbacher_2015b}.
Solar observations show that the flare-loop size does not necessarily scale in proportion to the flare energy \citep[e.g.,][]{Nagashima_2006} and there is no correlation for coronal active regions between magnetic flux density and loop size \citep[e.g.,][]{Yashiro_2001}.
Then, we refer to the map with $h \sim 0.1\,R_{\odot}$ and $\log(h/R_{*})=-2$ for these two systems.

For the flare detected on HR~1099, we estimate the flare-loop size using the magnetic-reconnection model of \citet{Shibata_2002},
\begin{eqnarray} 
l_{\mathrm{SY}} = &10^{9}& \left( \frac{EM_{\mathrm{peak}}}{10^{48} \: \mathrm{cm}^{-3}} \right)^{3/5} \nonumber \\ &\times& \left( \frac{\mathnormal{n_{0}}}{10^{9} \: \mathrm{cm}^{-3}} \right)^{-2/5} \left( \frac{\mathnormal{T_{\mathrm{peak}}}}{10^{7} \: \mathrm{K}} \right)^{-8/5}
\: \mathrm{cm}, \label{eq:Shibata_Yokoyama_L}
\end{eqnarray}
where $EM_{\mathrm{peak}}$ is the volume emission measure at the flare peak, $T_{\mathrm{peak}}$ is the peak electron temperature, and $n_{0}$ is the pre-flare coronal density.
We adopt the emission measure ($1.0\times10^{54}$~cm$^{-3}$) and temperature ($4.76$~keV) of the high-temperature thermal component obtained from the broad-band spectral modeling in \citet{Kurihara_2025b} as $EM_{\mathrm{peak}}$ and $T_{\mathrm{peak}}$ in Equation~\ref{eq:Shibata_Yokoyama_L}, and we assume a quiescent coronal density of $n_{0}=10^{10}~\mathrm{cm^{-3}}$ for HR~1099 \citep{Ness_2002}.
This yields $l_{\mathrm{SY}}\simeq 1.5\,R_{\odot}$ and a loop height of $h \sim l_{\mathrm{SY}}/\pi \sim 0.5\,R_{\odot}$, which is about an order of magnitude smaller than the stellar radius of HR~1099 \citep[$3.7\,R_{\odot}$;][]{Donati_1999}.
We therefore refer to the map with $\log(h/R_{*})=-1$ for this case.

Figure~\ref{fig:Mollwide_location}a highlights the region in the $\log(h/R_{*})=-2.0$ map (Figure~\ref{fig:Mollweide_map}a) that matches the 90\% confidence range of the GT~Mus equivalent width (Table~\ref{tab:best-fit}).
During the XRISM observation in August 2024, the CBP is inferred to have been located at longitudes $55^{\circ} \leq |\alpha|\leq 80^{\circ}$ inside the limb (Figure~\ref{fig:Mollwide_location}a).

Figure~\ref{fig:Mollwide_location}b shows the corresponding constraint for $\sigma$~Gem, for which no Fe~K$\alpha$ line was detected (Figure~\ref{fig:FeKa}b).
Even without a detection, the Resolve energy resolution enables a stringent upper limit of 3.9~eV on the equivalent width, which restricts the CBP location during the XRISM observation in November 2025 to just inside the limb at $85^{\circ}\leq |\alpha|\leq95^{\circ}$ (Figure~\ref{fig:Mollwide_location}b).

Figure~\ref{fig:Mollwide_location}c indicates that the flare loop in HR~1099 is inferred to have been located at $70^{\circ}\leq |\alpha|\leq 90^{\circ}$ inside the limb.
Finally, we note that the latitudes and longitudes shown in Figure~\ref{fig:Mollwide_location} do not directly correspond to stellar coordinates because the rotation axis is inclined with respect to the line of sight \citep{Inoue_2026}.

We also note that $h/R_{*}$ and the source-location coordinates are intrinsically degenerate, and $h$ must therefore be constrained independently, as illustrated above.
In many cases, however, this is unlikely to be a serious limitation.
For stellar flares, $h$ corresponds to the loop height, which can be estimated from magnetic reconnection models \citep{Shibata_1996, Yokoyama_1998, Shibata_1999, Shibata_2002, Shibata_2011} or from the temporal evolution of the flare plasma \citep{Reale_2007}.
For X-ray binaries, $h$ corresponds to the binary separation and can be inferred from the orbital parameters.

\begin{figure}[t]
\begin{center}
\includegraphics[width=0.49\textwidth]{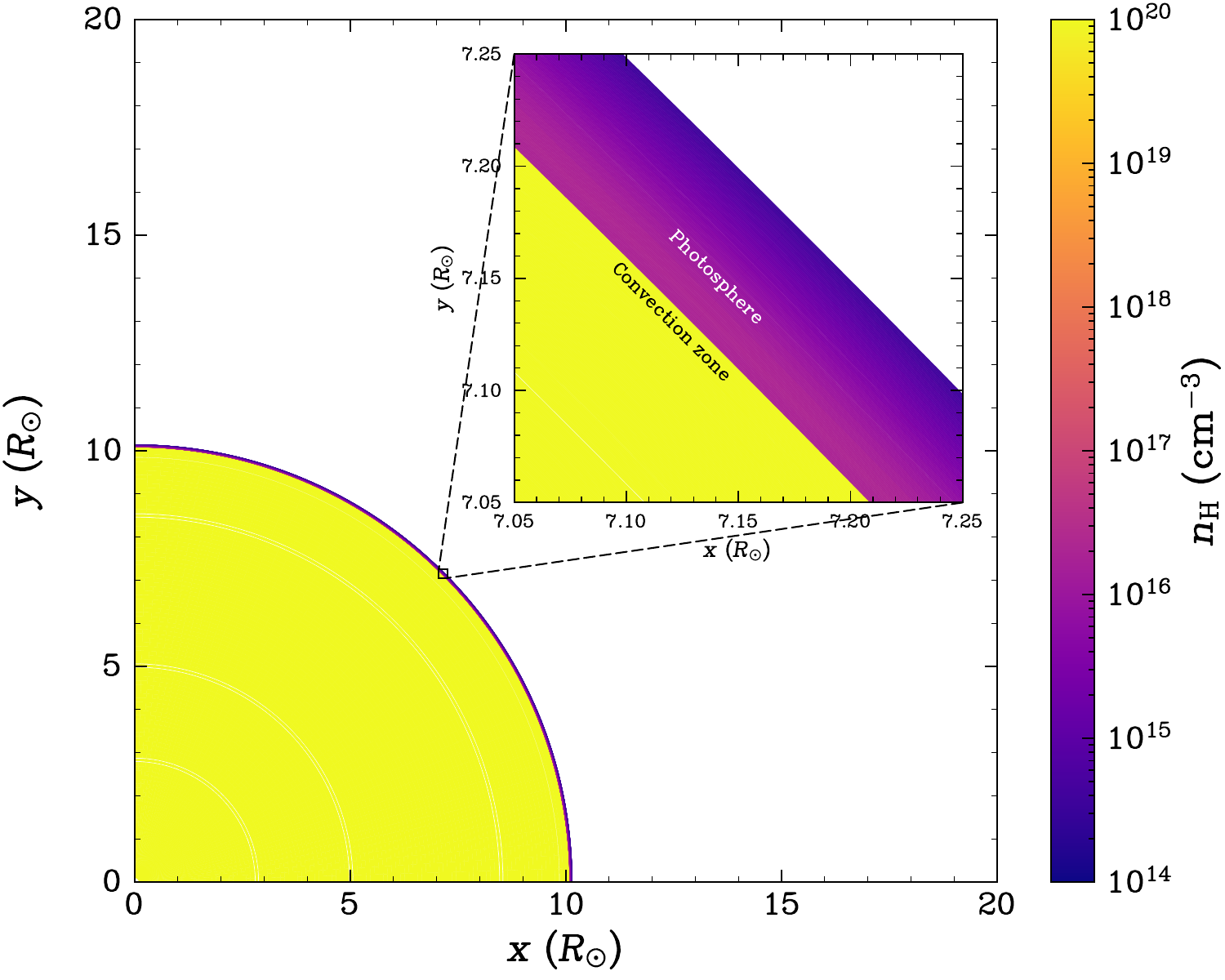}
\caption{
Cross-sectional view in the $x$--$y$ plane of the stellar density distribution input to \texttt{SKIRT}, based on the \texttt{PHOENIX} density structure (Figure \ref{fig:phoenix}b). The radial grid is divided into 100 bins over $0<r<10.05\,R_{\odot}$ and into 1000 bins over $10.05\,R_{\odot}<r<10.16\,R_{\odot}$. The polar and azimuthal angles are each divided into 100 bins over $0<\theta<\pi$ and $0<\phi<2\pi$, respectively.
{Alt text: Cross-sectional color map showing the stellar density distribution used as input to SKIRT.}
\label{fig:realistic_atmosphere}}
 \end{center}
\end{figure}

\begin{figure}[]
\begin{center}
\includegraphics[width=0.49\textwidth]{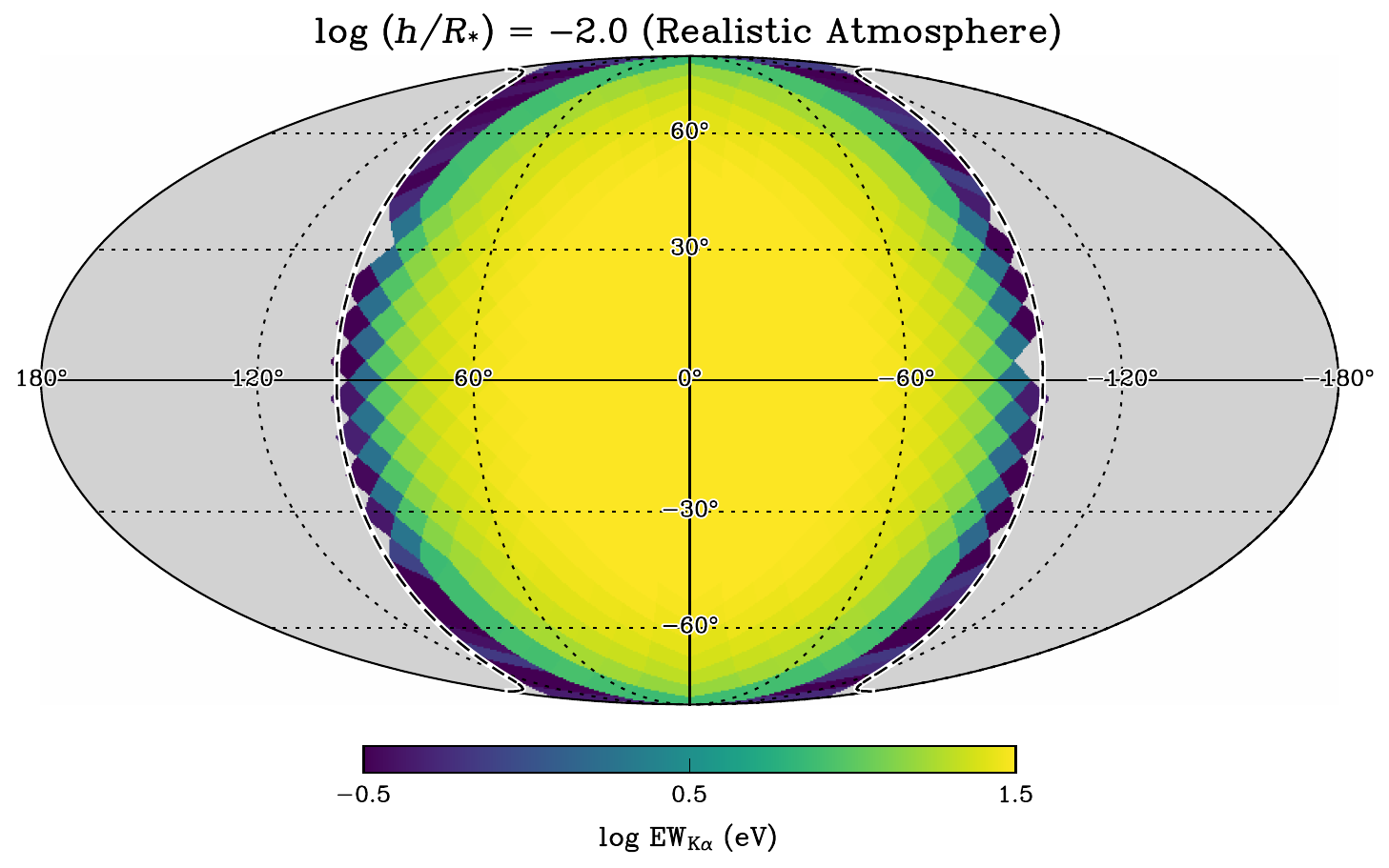}
\caption{
Equivalent-width map for $\log (h/R_{*}) = -2$ obtained from radiative-transfer calculations incorporating the atmospheric density and temperature structures (Figure \ref{fig:phoenix} and \ref{fig:realistic_atmosphere}).
{Alt text: Color map showing the Fe K-alpha equivalent-width map, calculated with atmospheric density and temperature structures included.}
}
\label{fig:realistic_results}
 \end{center}
\end{figure}

\subsection{Calculation with more realistic atmosphere model}\label{subsec:realistic}
When the emission source is located near or outside the limb, a more rigorous treatment would require accounting for the density structure of the atmospheric surface layers, because some photons reach the observer after skimming through the stellar surface layers.
As illustrated by the Sun, flares and active regions near the limb have greatly contributed to the understanding to their physical properties, making them especially important events  \citep[e.g., ][]{Tsuneta_1984, Tsuneta_1992, Uchida_1992, Masuda_1994, Tsuneta_1996, Tsuneta_1996b, Yokoyama_2001, Tomczak_2001, Krucker_2007, Krucker_2008, ODwyer_2011}.
Our map suggests that the active region was located near the limb when XRISM observed $\sigma$~Gem (Figure \ref{fig:Mollwide_location}b).
We therefore use $\sigma$~Gem as a case study to present an example of an application-specific calculation, incorporating an atmospheric density and temperature structure.

Figure~\ref{fig:phoenix} shows the atmospheric model computed with \texttt{PHOENIX} \citep{Husser_2013} for an effective temperature of 4600~K and a surface gravity of $g=10^{2.5}$~cm~s$^{-2}$, corresponding to the stellar parameters of $\sigma$~Gem \citep{Roettenbacher_2015}.
The \texttt{PHOENIX} output provides the temperature ($T$), pressure ($P$), and mass density ($\rho$) as functions of optical depth. For cool stars ($T_{\mathrm{eff}}<5000$~K), this optical depth corresponds to $\tau_{12000\,\mathrm{\mathring{A}}}$, defined at a wavelength of 12000~\AA\ \citep{Husser_2013}.

Assuming hydrostatic equilibrium, we treated the radius at which $\tau_{12000 \mathrm{\mathring{A}}}=2/3$ as corresponding to the observationally determined stellar radius \citep[$R_{*} = 10.1 R_{\odot}$; ][]{Roettenbacher_2015} and converted the optical depth $\tau_{12000 \mathrm{\mathring{A}}}=2/3$ to the radial distance $r$ from the stellar center (Figure~\ref{fig:phoenix}a).
Hydrostatic equilibrium gives
\begin{equation}
    \frac{dP}{dz}=-\rho g,
\end{equation}
where $z$ is the geometric depth measured from the $\tau_{12000\,\mathrm{\mathring{A}}}=2/3$ surface ($z=0$) and increases toward the stellar center.
The radial coordinate is then obtained as
\begin{equation}
    r = R_{*}-z = R_{*} + \int \frac{dP}{\rho g}.
\end{equation}
We also convert the mass density to the hydrogen number density (Figure~\ref{fig:phoenix}b) via
\begin{equation}
    n_{\mathrm{H}} = \frac{\rho X}{m_{\mathrm{H}}},
\end{equation}
where $X=0.738$ is the solar hydrogen mass fraction \citep{Asplund_2009} and $m_{\mathrm{H}}=1.7\times10^{-24}$~g is the hydrogen mass.
Finally, following Equation~(1) of \citet{Nagai_2026}, we compute the Fe K edge optical depth (Figure~\ref{fig:phoenix}d) as
\begin{equation}
    \tau_{\mathrm{FeK}}(z)=A_{\mathrm{Fe}}\,\sigma_{\mathrm{FeK}} \int_{z_{0}}^{z} n_{\mathrm{H}}(z)\,dz,
\end{equation}
where $A_{\mathrm{Fe}}=6.7\times10^{-5}$ is the solar Fe abundance relative to H \citep{Wilms_2000}, $\sigma_{\mathrm{FeK}}=3.8\times10^{-20}$~cm$^{2}$ is the photoelectric absorption cross section at the Fe K edge for neutral Fe \citep{Berger_2010}, and $z_{0}$ is the geometrical depth where $\tau_{12000\,\mathrm{\mathring{A}}}=0$.

Figure \ref{fig:realistic_atmosphere} shows the stellar density distribution input to \texttt{SKIRT}, constructed from the \texttt{PHOENIX} data and the parameter conversions described above.
We define $R_{\mathrm{FeK}}$ as the radius where $\tau_{\mathrm{FeK}}=2/3$ and $R_{0}$ as the radius where $\tau_{12000\,\mathrm{\mathring{A}}}=100$. 
We adopt the density and temperature structure (Figure \ref{fig:phoenix}b \& c) in the layer $R_{0}<r<R_{\mathrm{FeK}}$, and assume a constant hydrogen number density of $n_{\mathrm{H}}=10^{20}$~cm$^{-3}$ and temperature of $10^{5}$ K for $r<R_{0}$, corresponding to layers below the convective zone.
We specified the spatial grid in spherical coordinates with \texttt{Sphere3DSpatialGrid} class in \texttt{SKIRT}. To reduce the computational cost, we adopted a relatively coarse radial grid with 100 bins in the stellar interior ($0<r<10.05\,R_{\odot}$), while we used a finer grid with 1000 bins near the surface ($10.05\,R_{\odot}<r<10.16\,R_{\odot}$) to ensure that the atmospheric density structure is adequately resolved in the simulations.
All other settings were kept identical to those described in Section \ref{subsec:setup}, and we performed the calculations for both cases in which the emission source is placed inside and outside the limb.
When the continuum flux at 6.3 keV fell below 10$^{-10}$ photons keV$^{-1}$ s$^{-1}$ cm$^{-2}$, we treated the Fe K$\alpha$ equivalent width as undefined.

Figure \ref{fig:realistic_results} shows the radiative transfer calculation results with the realistic atmosphere described above.
The resulting map is essentially unchanged from that computed with a uniform-density sphere (Figure \ref{fig:Mollweide_map}a), both inside and outside the limb.
In principle, when the surface density structure is included, some photons should graze the stellar surface and still reach the observer just outside the limb, so the Fe~K$\alpha$ EW should remain definable there. However, this effect is not captured with our current spherical tessellation ($N_{\mathrm{side}} = 9$).
The key result here is that the map inside the limb is insensitive to whether the atmospheric density structure is included. This implies that the maps shown in Figure \ref{fig:Mollweide_map} should be applicable to stars with different surface-atmosphere structures.

\section{Conclusion}\label{sec:conclusion}
\label{sec:conclusion}
With the advent of the microcalorimeter Resolve, the detectable Fe K$\alpha$ equivalent width has been pushed down from $\sim 50$ eV to $\sim 5$ eV.
Even for equivalent widths large enough to have been detectable with previous missions, the measurement precision has improved, reducing the uncertainty range by more than an order of magnitude.
Moreover, when no Fe K$\alpha$ line is detected, stringent upper limits as low as a few eV can be obtained, making non-detections themselves valuable for constraining the geometry. 
Motivated by this microcalorimeter era, we performed three-dimensional radiative-transfer calculations with \texttt{SKIRT} and constructed generalized maps of the Fe K$\alpha$ equivalent width for a system consisting of a point source and a spherical reflector.
By comparing the Resolve spectra of RS CVn-type stars with our maps, we constrained the locations of flares and CBPs, thereby demonstrating the utility of the maps.

Although our maps carry uncertainties of a factor of a few due to differences for the input SED, they are broadly applicable to systems consisting of a point source and a spherical reflector, such as stellar flares with a flare loop and a photosphere, X-ray binaries with a neutron star and a main-sequence star, and cataclysmic variables with a white dwarf and a main-sequence star. 
By clarifying the geometric picture of these systems, our maps should also contribute substantially to their physical understanding.
For example, in the case of stellar flares and active regions, determining their latitude and longitude distribution with our maps would enable us to discuss the stellar global magnetic-field configuration \citep[e.g.,][]{Nandy_2002, Hathaway_2010, Charbonneau_2020, Yang_2026}.

\begin{ack} 
This research has made use of data and/or software provided by the High Energy Astrophysics Science Archive Research Center (HEASARC), which is a service of the Astrophysics Science Division at NASA/GSFC.
In this research work we used the supercomputer of ACCMS, Kyoto University.
We thank Yuto Mochizuki (ISAS/JAXA) and Yutaro Nagai (Kyoto University) for their useful discussions.
This research is supported by JSPS Core-to-Core Program (grant number: JPJSCCA20220002) and JSPS KAKENHI grant No. 24KJ1483 (S.I.).
T.E. was supported by the JST, Japan grant number JPMJFR202O (Sohatsu) and JSPS KAKENHI Grant Number 26H02075.
\end{ack}


\bibliography{reference}

@ARTICLE{Tsuru_1989,
       author = {{Tsuru}, T. and {Makishima}, K. and {Ohashi}, T. and {Inoue}, H. and {Koyama}, K. and {Turner}, M.~J.~L. and {Barstow}, M.~A. and {McHardy}, I.~M. and {Pye}, J.~P. and {Tsunemi}, H. and {Kitamoto}, S. and {Taylor}, A.~R. and {Nelson}, R.~F.},
        title = "{X-ray and radio observations of flares from the RS Canum Venaticorum system UX Arietis.}",
      journal = {\pasj},
         year = 1989,
        month = jan,
       volume = {41},
        pages = {679-695},
       adsurl = {https://ui.adsabs.harvard.edu/abs/1989PASJ...41..679T}
}

@ARTICLE{Tsuboi_2016,
       author = {{Tsuboi}, Yohko and {Yamazaki}, Kyohei and {Sugawara}, Yasuharu and {Kawagoe}, Atsushi and {Kaneto}, Soichiro and {Iizuka}, Ryo and {Matsumura}, Takanori and {Nakahira}, Satoshi and {Higa}, Masaya and {Matsuoka}, Masaru and {Sugizaki}, Mutsumi and {Ueda}, Yoshihiro and {Kawai}, Nobuyuki and {Morii}, Mikio and {Serino}, Motoko and {Mihara}, Tatehiro and {Tomida}, Hiroshi and {Ueno}, Shiro and {Negoro}, Hitoshi and {Daikyuji}, Arata and {Ebisawa}, Ken and {Eguchi}, Satoshi and {Hiroi}, Kazuo and {Ishikawa}, Masaki and {Isobe}, Naoki and {Kawasaki}, Kazuyoshi and {Kimura}, Masashi and {Kitayama}, Hiroki and {Kohama}, Mitsuhiro and {Kotani}, Taro and {Nakagawa}, Yujin E. and {Nakajima}, Motoki and {Ozawa}, Hiroshi and {Shidatsu}, Megumi and {Sootome}, Tetsuya and {Sugimori}, Kousuke and {Suwa}, Fumitoshi and {Tsunemi}, Hiroshi and {Usui}, Ryuichi and {Yamamoto}, Takayuki and {Yamaoka}, Kazutaka and {Yoshida}, Atsumasa},
        title = "{Large X-ray flares on stars detected with MAXI/GSC: A universal correlation between the duration of a flare and its X-ray luminosity}",
      journal = {\pasj},
         year = 2016,
        month = oct,
       volume = {68},
       number = {5},
          eid = {90},
        pages = {90},
          doi = {10.1093/pasj/psw081},
archivePrefix = {arXiv},
       eprint = {1609.01925},
 primaryClass = {astro-ph.HE},
       adsurl = {https://ui.adsabs.harvard.edu/abs/2016PASJ...68...90T}
}

@ARTICLE{Kurihara_2024,
       author = {{Kurihara}, Miki and {Iwakiri}, Wataru Buz and {Tsujimoto}, Masahiro and {Ebisawa}, Ken and {Toriumi}, Shin and {Imada}, Shinsuke and {Tsuboi}, Yohko and {Usui}, Kazuki and {Gendreau}, Keith C. and {Arzoumanian}, Zaven},
        title = "{Investigation of Nonequilibrium Ionization Plasma during a Giant Flare of UX Arietis Triggered with MAXI and Observed with NICER}",
      journal = {\apj},
         year = 2024,
        month = apr,
       volume = {965},
       number = {2},
          eid = {135},
        pages = {135},
          doi = {10.3847/1538-4357/ad35c5},
archivePrefix = {arXiv},
       eprint = {2403.12351},
 primaryClass = {astro-ph.SR},
       adsurl = {https://ui.adsabs.harvard.edu/abs/2024ApJ...965..135K}
}

@ARTICLE{Sasaki_2021,
       author = {{Sasaki}, Ryo and {Tsuboi}, Yohko and {Iwakiri}, Wataru and {Nakahira}, Satoshi and {Maeda}, Yoshitomo and {Gendreau}, Keith and {Corcoran}, Michael F. and {Hamaguchi}, Kenji and {Arzoumanian}, Zaven and {Markwardt}, Craig B. and {Enoto}, Teruaki and {Sato}, Tatsuki and {Kawai}, Hiroki and {Mihara}, Tatehiro and {Shidatsu}, Megumi and {Negoro}, Hitoshi and {Serino}, Motoko},
        title = "{The RS CVn-type Star GT Mus Shows Most Energetic X-Ray Flares Throughout the 2010s}",
      journal = {\apj},
         year = 2021,
        month = mar,
       volume = {910},
       number = {1},
          eid = {25},
        pages = {25},
          doi = {10.3847/1538-4357/abde38},
archivePrefix = {arXiv},
       eprint = {2103.16822},
 primaryClass = {astro-ph.HE},
       adsurl = {https://ui.adsabs.harvard.edu/abs/2021ApJ...910...25S}
}

@ARTICLE{Canizares_2005,
       author = {{Canizares}, Claude R. and {Davis}, John E. and {Dewey}, Daniel and {Flanagan}, Kathryn A. and {Galton}, Eugene B. and {Huenemoerder}, David P. and {Ishibashi}, Kazunori and {Markert}, Thomas H. and {Marshall}, Herman L. and {McGuirk}, Michael and {Schattenburg}, Mark L. and {Schulz}, Norbert S. and {Smith}, Henry I. and {Wise}, Michael},
        title = "{The Chandra High-Energy Transmission Grating: Design, Fabrication, Ground Calibration, and 5 Years in Flight}",
      journal = {\pasp},
         year = 2005,
        month = oct,
       volume = {117},
       number = {836},
        pages = {1144-1171},
          doi = {10.1086/432898},
archivePrefix = {arXiv},
       eprint = {astro-ph/0507035},
 primaryClass = {astro-ph},
       adsurl = {https://ui.adsabs.harvard.edu/abs/2005PASP..117.1144C}
}

@ARTICLE{Smith_2001,
       author = {{Smith}, Randall K. and {Brickhouse}, Nancy S. and {Liedahl}, Duane A. and {Raymond}, John C.},
        title = "{Collisional Plasma Models with APEC/APED: Emission-Line Diagnostics of Hydrogen-like and Helium-like Ions}",
      journal = {\apjl},
         year = 2001,
        month = aug,
       volume = {556},
       number = {2},
        pages = {L91-L95},
          doi = {10.1086/322992},
archivePrefix = {arXiv},
       eprint = {astro-ph/0106478},
 primaryClass = {astro-ph},
       adsurl = {https://ui.adsabs.harvard.edu/abs/2001ApJ...556L..91S}
}

@ARTICLE{Shibata_2002,
       author = {{Shibata}, Kazunari and {Yokoyama}, Takaaki},
        title = "{A Hertzsprung-Russell-like Diagram for Solar/Stellar Flares and Corona: Emission Measure versus Temperature Diagram}",
      journal = {\apj},
         year = 2002,
        month = sep,
       volume = {577},
       number = {1},
        pages = {422-432},
          doi = {10.1086/342141},
archivePrefix = {arXiv},
       eprint = {astro-ph/0206016},
 primaryClass = {astro-ph},
       adsurl = {https://ui.adsabs.harvard.edu/abs/2002ApJ...577..422S}
}

@ARTICLE{Reale_2007,
       author = {{Reale}, F.},
        title = "{Diagnostics of stellar flares from X-ray observations: from the decay to the rise phase}",
      journal = {\aap},
         year = 2007,
        month = aug,
       volume = {471},
       number = {1},
        pages = {271-279},
          doi = {10.1051/0004-6361:20077223},
archivePrefix = {arXiv},
       eprint = {0705.3254},
 primaryClass = {astro-ph},
       adsurl = {https://ui.adsabs.harvard.edu/abs/2007A&A...471..271R}
}

@ARTICLE{Inoue_2024b,
       author = {{Inoue}, Shun and {Iwakiri}, Wataru Buz and {Enoto}, Teruaki and {Uchida}, Hiroyuki and {Kurihara}, Miki and {Tsujimoto}, Masahiro and {Notsu}, Yuta and {Hamaguchi}, Kenji and {Gendreau}, Keith and {Arzoumanian}, Zaven and {Tsuru}, Takeshi Go},
        title = "{High-velocity Blue-shifted Fe XXV He{\ensuremath{\alpha}} Line during a Superflare of the RS Canum Venaticorum{\textendash}type Star IM Peg}",
      journal = {\apjl},
         year = "2024b",
        month = jul,
       volume = {969},
       number = {1},
          eid = {L12},
        pages = {L12},
          doi = {10.3847/2041-8213/ad5667},
       adsurl = {https://ui.adsabs.harvard.edu/abs/2024ApJ...969L..12I}
}

@ARTICLE{Osten_2007,
       author = {{Osten}, Rachel A. and {Drake}, Stephen and {Tueller}, Jack and {Cummings}, Jay and {Perri}, Matteo and {Moretti}, Alberto and {Covino}, Stefano},
        title = "{Nonthermal Hard X-Ray Emission and Iron K{\ensuremath{\alpha}} Emission from a Superflare on II Pegasi}",
      journal = {\apj},
         year = 2007,
        month = jan,
       volume = {654},
       number = {2},
        pages = {1052-1067},
          doi = {10.1086/509252},
archivePrefix = {arXiv},
       eprint = {astro-ph/0609205},
 primaryClass = {astro-ph},
       adsurl = {https://ui.adsabs.harvard.edu/abs/2007ApJ...654.1052O}
}

@ARTICLE{Osten_2010,
       author = {{Osten}, Rachel A. and {Godet}, Olivier and {Drake}, Stephen and {Tueller}, Jack and {Cummings}, Jay and {Krimm}, Hans and {Pye}, John and {Pal'shin}, Valentin and {Golenetskii}, Sergei and {Reale}, Fabio and {Oates}, Samantha R. and {Page}, Mat J. and {Melandri}, Andrea},
        title = "{The Mouse That Roared: A Superflare from the dMe Flare Star EV Lac Detected by Swift and Konus-Wind}",
      journal = {\apj},
         year = 2010,
        month = sep,
       volume = {721},
       number = {1},
        pages = {785-801},
          doi = {10.1088/0004-637X/721/1/785},
archivePrefix = {arXiv},
       eprint = {1007.5300},
 primaryClass = {astro-ph.SR},
       adsurl = {https://ui.adsabs.harvard.edu/abs/2010ApJ...721..785O}
}

@ARTICLE{Karmakar_2017,
       author = {{Karmakar}, Subhajeet and {Pandey}, J.~C. and {Airapetian}, V.~S. and {Misra}, Kuntal},
        title = "{X-Ray Superflares on CC Eri}",
      journal = {\apj},
         year = 2017,
        month = may,
       volume = {840},
       number = {2},
          eid = {102},
        pages = {102},
          doi = {10.3847/1538-4357/aa6cb0},
archivePrefix = {arXiv},
       eprint = {1705.06930},
 primaryClass = {astro-ph.SR},
       adsurl = {https://ui.adsabs.harvard.edu/abs/2017ApJ...840..102K}
}

@ARTICLE{Testa_2008,
       author = {{Testa}, Paola and {Drake}, Jeremy J. and {Ercolano}, Barbara and {Reale}, Fabio and {Huenemoerder}, David P. and {Affer}, Laura and {Micela}, Giuseppina and {Garcia-Alvarez}, David},
        title = "{Geometry Diagnostics of a Stellar Flare from Fluorescent X-Rays}",
      journal = {\apjl},
         year = 2008,
        month = mar,
       volume = {675},
       number = {2},
        pages = {L97},
          doi = {10.1086/533461},
archivePrefix = {arXiv},
       eprint = {0801.3857},
 primaryClass = {astro-ph},
       adsurl = {https://ui.adsabs.harvard.edu/abs/2008ApJ...675L..97T}
}

@ARTICLE{Huenemoerder_2010,
       author = {{Huenemoerder}, David P. and {Schulz}, Norbert S. and {Testa}, Paola and {Drake}, Jeremy J. and {Osten}, Rachel A. and {Reale}, Fabio},
        title = "{X-ray Flares of EV Lac: Statistics, Spectra, and Diagnostics}",
      journal = {\apj},
         year = 2010,
        month = nov,
       volume = {723},
       number = {2},
        pages = {1558-1567},
          doi = {10.1088/0004-637X/723/2/1558},
archivePrefix = {arXiv},
       eprint = {1006.2558},
 primaryClass = {astro-ph.HE},
       adsurl = {https://ui.adsabs.harvard.edu/abs/2010ApJ...723.1558H}
}

@ARTICLE{Drake_2008,
       author = {{Drake}, Jeremy J. and {Ercolano}, Barbara and {Swartz}, Douglas A.},
        title = "{X-Ray-fluorescent Fe K{\ensuremath{\alpha}} Lines from Stellar Photospheres}",
      journal = {\apj},
         year = 2008,
        month = may,
       volume = {678},
       number = {1},
        pages = {385-393},
          doi = {10.1086/524976},
archivePrefix = {arXiv},
       eprint = {0710.0621},
 primaryClass = {astro-ph},
       adsurl = {https://ui.adsabs.harvard.edu/abs/2008ApJ...678..385D}
}

@ARTICLE{Zarro_1992,
       author = {{Zarro}, D.~M. and {Dennis}, B.~R. and {Slater}, G.~L.},
        title = "{Impulsive Phase Fe K alpha Emission in a Flare of 1989 March}",
      journal = {\apj},
         year = 1992,
        month = jun,
       volume = {391},
        pages = {865},
          doi = {10.1086/171395},
       adsurl = {https://ui.adsabs.harvard.edu/abs/1992ApJ...391..865Z}
}

@ARTICLE{Tanaka_1984,
       author = {{Tanaka}, K. and {Watanabe}, T. and {Nitta}, N.},
        title = "{Solar flare iron K-alpha emission associated with a hard X-ray burst}",
      journal = {\apj},
         year = 1984,
        month = jul,
       volume = {282},
        pages = {793-798},
          doi = {10.1086/162264},
       adsurl = {https://ui.adsabs.harvard.edu/abs/1984ApJ...282..793T}
}

@ARTICLE{Emslie_1986,
       author = {{Emslie}, A.~G. and {Phillips}, K.~J.~H. and {Dennis}, B.~R.},
        title = "{The excitation of the iron K-alpha feature in solar flares}",
      journal = {\solphys},
         year = 1986,
        month = jan,
       volume = {103},
        pages = {89-102},
       adsurl = {https://ui.adsabs.harvard.edu/abs/1986SoPh..103...89E}
}

@ARTICLE{Parmar_1984,
       author = {{Parmar}, A.~N. and {Culhane}, J.~L. and {Rapley}, C.~G. and {Wolfson}, C.~J. and {Acton}, L.~W. and {Phillips}, K.~J.~H. and {Dennis}, B.~R.},
        title = "{SMM observations of K-alpha radiation from fluorescence of photospheric iron by solar flare X-rays}",
      journal = {\apj},
         year = 1984,
        month = apr,
       volume = {279},
        pages = {866-874},
          doi = {10.1086/161957},
       adsurl = {https://ui.adsabs.harvard.edu/abs/1984ApJ...279..866P}
}

@ARTICLE{Bai_1979,
       author = {{Bai}, T.},
        title = "{Iron K{\ensuremath{\alpha}}-fluorescence in solar flares: A probe of the photospheric iron abundance}",
      journal = {\solphys},
         year = 1979,
        month = may,
       volume = {62},
       number = {1},
        pages = {113-121},
          doi = {10.1007/BF00150138},
       adsurl = {https://ui.adsabs.harvard.edu/abs/1979SoPh...62..113B}
}

@ARTICLE{Neupert_1967,
       author = {{Neupert}, W.~M. and {Gates}, W. and {Swartz}, M. and {Young}, R.},
        title = "{Observation of the Solar Flare X-Ray Emission-Line Spectrum of Iron from 1.3 to 20 {\r{A}}}",
      journal = {\apjl},
         year = 1967,
        month = aug,
       volume = {149},
        pages = {L79},
          doi = {10.1086/180061},
       adsurl = {https://ui.adsabs.harvard.edu/abs/1967ApJ...149L..79N}
}

@ARTICLE{Neupert_1971,
       author = {{Neupert}, Werner M.},
        title = "{Satellite lines in the solar X-ray spectrum.}",
      journal = {\solphys},
         year = 1971,
        month = jul,
       volume = {18},
       number = {3},
        pages = {474-488},
          doi = {10.1007/BF00149069},
       adsurl = {https://ui.adsabs.harvard.edu/abs/1971SoPh...18..474N}
}

@ARTICLE{Doschek_1971,
       author = {{Doschek}, G.~A. and {Meekins}, J.~F. and {Kreplin}, R.~W. and {Chubb}, T.~A. and {Friedman}, H.},
        title = "{Iron-Line Emission during Solar Flares}",
      journal = {\apj},
         year = 1971,
        month = dec,
       volume = {170},
        pages = {573},
          doi = {10.1086/151243},
       adsurl = {https://ui.adsabs.harvard.edu/abs/1971ApJ...170..573D}
}

@ARTICLE{Feldman_1980,
       author = {{Feldman}, U. and {Doschek}, G.~A. and {Kreplin}, R.~W.},
        title = "{High-resolution X-ray spectra of the 1979 March 25 solar flare}",
      journal = {\apj},
         year = 1980,
        month = may,
       volume = {238},
        pages = {365-374},
          doi = {10.1086/157993},
       adsurl = {https://ui.adsabs.harvard.edu/abs/1980ApJ...238..365F}
}

@ARTICLE{Culhane_1981,
       author = {{Culhane}, J.~L. and {Rapley}, C.~G. and {Bentley}, R.~D. and {Gabriel}, A.~H. and {Phillips}, K.~J. and {Acton}, L.~W. and {Wolfson}, C.~J. and {Catura}, R.~C. and {Jordan}, C. and {Antonucci}, E.},
        title = "{X-ray spectra of solar flares obtained with a high-resolution bent crystal spectrometer}",
      journal = {\apjl},
         year = 1981,
        month = mar,
       volume = {244},
        pages = {L141-L145},
          doi = {10.1086/183499},
       adsurl = {https://ui.adsabs.harvard.edu/abs/1981ApJ...244L.141C}
}

@ARTICLE{Doyle_1991,
       author = {{Doyle}, J.~G. and {Kellett}, B.~J. and {Byrne}, P.~B. and {Avgoloupis}, S. and {Mavridis}, L.~N. and {Seiradakis}, J.~H. and {Bromage}, G.~E. and {Tsuru}, T. and {Makishima}, K. and {Makishima}, K. and {McHardy}, I.~M.},
        title = "{Simultaneous detection of a large flare in the X-ray and optical regions on the RS CVn-type star II Peg.}",
      journal = {\mnras},
         year = 1991,
        month = feb,
       volume = {248},
        pages = {503},
          doi = {10.1093/mnras/248.3.503},
       adsurl = {https://ui.adsabs.harvard.edu/abs/1991MNRAS.248..503D}
}

@ARTICLE{Osten_2000,
       author = {{Osten}, Rachel A. and {Brown}, Alexander and {Ayres}, Thomas R. and {Linsky}, Jeffrey L. and {Drake}, Stephen A. and {Gagn{\'e}}, Marc and {Stern}, Robert A.},
        title = "{Radio, X-Ray, and Extreme-Ultraviolet Coronal Variability of the Short-Period RS Canum Venaticorum Binary {\ensuremath{\sigma}}$^{2}$ Coronae Borealis}",
      journal = {\apj},
         year = 2000,
        month = dec,
       volume = {544},
       number = {2},
        pages = {953-976},
          doi = {10.1086/317249},
       adsurl = {https://ui.adsabs.harvard.edu/abs/2000ApJ...544..953O}
}

@ARTICLE{Ercolano_2008,
       author = {{Ercolano}, Barbara and {Drake}, Jeremy J. and {Reale}, Fabio and {Testa}, Paola and {Miller}, Jon M.},
        title = "{Fe K{\ensuremath{\alpha}} and Hydrodynamic Loop Model Diagnostics for a Large Flare on II Pegasi}",
      journal = {\apj},
         year = 2008,
        month = dec,
       volume = {688},
       number = {2},
        pages = {1315-1319},
          doi = {10.1086/591934},
archivePrefix = {arXiv},
       eprint = {0807.2093},
 primaryClass = {astro-ph},
       adsurl = {https://ui.adsabs.harvard.edu/abs/2008ApJ...688.1315E}
}

@ARTICLE{Tanaka_1985,
       author = {{Tanaka}, K. and {Zirin}, H.},
        title = "{The great flare of 1982 June 6}",
      journal = {\apj},
         year = 1985,
        month = dec,
       volume = {299},
        pages = {1036-1046},
          doi = {10.1086/163771},
       adsurl = {https://ui.adsabs.harvard.edu/abs/1985ApJ...299.1036T}
}

@ARTICLE{Roettenbacher_2015,
       author = {{Roettenbacher}, Rachael M. and {Monnier}, John D. and {Henry}, Gregory W. and {Fekel}, Francis C. and {Williamson}, Michael H. and {Pourbaix}, Dimitri and {Latham}, David W. and {Latham}, Christian A. and {Torres}, Guillermo and {Baron}, Fabien and {Che}, Xiao and {Kraus}, Stefan and {Schaefer}, Gail H. and {Aarnio}, Alicia N. and {Korhonen}, Heidi and {Harmon}, Robert O. and {ten Brummelaar}, Theo A. and {Sturmann}, Judit and {Sturmann}, Laszlo and {Turner}, Nils H.},
        title = "{Detecting the Companions and Ellipsoidal Variations of RS CVn Primaries. I. {\ensuremath{\sigma}} Geminorum}",
      journal = {\apj},
         year = 2015,
        month = jul,
       volume = {807},
       number = {1},
          eid = {23},
        pages = {23},
          doi = {10.1088/0004-637X/807/1/23},
archivePrefix = {arXiv},
       eprint = {1504.06628},
 primaryClass = {astro-ph.SR},
       adsurl = {https://ui.adsabs.harvard.edu/abs/2015ApJ...807...23R}
}

@ARTICLE{Bommier_2020,
       author = {{Bommier}, V{\'e}ronique},
        title = "{Solar photosphere magnetization}",
      journal = {\aap},
         year = 2020,
        month = feb,
       volume = {634},
          eid = {A40},
        pages = {A40},
          doi = {10.1051/0004-6361/201935244},
archivePrefix = {arXiv},
       eprint = {1907.06476},
 primaryClass = {astro-ph.SR},
       adsurl = {https://ui.adsabs.harvard.edu/abs/2020A&A...634A..40B}
}

@ARTICLE{Gaia_2016,
       author = {{Gaia Collaboration} and {Prusti}, T. and {de Bruijne}, J.~H.~J. and {Brown}, A.~G.~A. and {Vallenari}, A. and {Babusiaux}, C. and {Bailer-Jones}, C.~A.~L. and {Bastian}, U. and {Biermann}, M. and {Evans}, D.~W. and {Eyer}, L. and {Jansen}, F. and {Jordi}, C. and {Klioner}, S.~A. and {Lammers}, U. and {Lindegren}, L. and {Luri}, X. and {Mignard}, F. and {Milligan}, D.~J. and {Panem}, C. and {Poinsignon}, V. and {Pourbaix}, D. and {Randich}, S. and {Sarri}, G. and {Sartoretti}, P. and {Siddiqui}, H.~I. and {Soubiran}, C. and {Valette}, V. and {van Leeuwen}, F. and {Walton}, N.~A. and {Aerts}, C. and {Arenou}, F. and {Cropper}, M. and {Drimmel}, R. and {H{\o}g}, E. and {Katz}, D. and {Lattanzi}, M.~G. and {O'Mullane}, W. and {Grebel}, E.~K. and {Holland}, A.~D. and {Huc}, C. and {Passot}, X. and {Bramante}, L. and {Cacciari}, C. and {Casta{\~n}eda}, J. and {Chaoul}, L. and {Cheek}, N. and {De Angeli}, F. and {Fabricius}, C. and {Guerra}, R. and {Hern{\'a}ndez}, J. and {Jean-Antoine-Piccolo}, A. and {Masana}, E. and {Messineo}, R. and {Mowlavi}, N. and {Nienartowicz}, K. and {Ord{\'o}{\~n}ez-Blanco}, D. and {Panuzzo}, P. and {Portell}, J. and {Richards}, P.~J. and {Riello}, M. and {Seabroke}, G.~M. and {Tanga}, P. and {Th{\'e}venin}, F. and {Torra}, J. and {Els}, S.~G. and {Gracia-Abril}, G. and {Comoretto}, G. and {Garcia-Reinaldos}, M. and {Lock}, T. and {Mercier}, E. and {Altmann}, M. and {Andrae}, R. and {Astraatmadja}, T.~L. and {Bellas-Velidis}, I. and {Benson}, K. and {Berthier}, J. and {Blomme}, R. and {Busso}, G. and {Carry}, B. and {Cellino}, A. and {Clementini}, G. and {Cowell}, S. and {Creevey}, O. and {Cuypers}, J. and {Davidson}, M. and {De Ridder}, J. and {de Torres}, A. and {Delchambre}, L. and {Dell'Oro}, A. and {Ducourant}, C. and {Fr{\'e}mat}, Y. and {Garc{\'\i}a-Torres}, M. and {Gosset}, E. and {Halbwachs}, J. -L. and {Hambly}, N.~C. and {Harrison}, D.~L. and {Hauser}, M. and {Hestroffer}, D. and {Hodgkin}, S.~T. and {Huckle}, H.~E. and {Hutton}, A. and {Jasniewicz}, G. and {Jordan}, S. and {Kontizas}, M. and {Korn}, A.~J. and {Lanzafame}, A.~C. and {Manteiga}, M. and {Moitinho}, A. and {Muinonen}, K. and {Osinde}, J. and {Pancino}, E. and {Pauwels}, T. and {Petit}, J. -M. and {Recio-Blanco}, A. and {Robin}, A.~C. and {Sarro}, L.~M. and {Siopis}, C. and {Smith}, M. and {Smith}, K.~W. and {Sozzetti}, A. and {Thuillot}, W. and {van Reeven}, W. and {Viala}, Y. and {Abbas}, U. and {Abreu Aramburu}, A. and {Accart}, S. and {Aguado}, J.~J. and {Allan}, P.~M. and {Allasia}, W. and {Altavilla}, G. and {{\'A}lvarez}, M.~A. and {Alves}, J. and {Anderson}, R.~I. and {Andrei}, A.~H. and {Anglada Varela}, E. and {Antiche}, E. and {Antoja}, T. and {Ant{\'o}n}, S. and {Arcay}, B. and {Atzei}, A. and {Ayache}, L. and {Bach}, N. and {Baker}, S.~G. and {Balaguer-N{\'u}{\~n}ez}, L. and {Barache}, C. and {Barata}, C. and {Barbier}, A. and {Barblan}, F. and {Baroni}, M. and {Barrado y Navascu{\'e}s}, D. and {Barros}, M. and {Barstow}, M.~A. and {Becciani}, U. and {Bellazzini}, M. and {Bellei}, G. and {Bello Garc{\'\i}a}, A. and {Belokurov}, V. and {Bendjoya}, P. and {Berihuete}, A. and {Bianchi}, L. and {Bienaym{\'e}}, O. and {Billebaud}, F. and {Blagorodnova}, N. and {Blanco-Cuaresma}, S. and {Boch}, T. and {Bombrun}, A. and {Borrachero}, R. and {Bouquillon}, S. and {Bourda}, G. and {Bouy}, H. and {Bragaglia}, A. and {Breddels}, M.~A. and {Brouillet}, N. and {Br{\"u}semeister}, T. and {Bucciarelli}, B. and {Budnik}, F. and {Burgess}, P. and {Burgon}, R. and {Burlacu}, A. and {Busonero}, D. and {Buzzi}, R. and {Caffau}, E. and {Cambras}, J. and {Campbell}, H. and {Cancelliere}, R. and {Cantat-Gaudin}, T. and {Carlucci}, T. and {Carrasco}, J.~M. and {Castellani}, M. and {Charlot}, P. and {Charnas}, J. and {Charvet}, P. and {Chassat}, F. and {Chiavassa}, A. and {Clotet}, M. and {Cocozza}, G. and {Collins}, R.~S. and {Collins}, P. and {Costigan}, G. and {Crifo}, F. and {Cross}, N.~J.~G. and {Crosta}, M. and {Crowley}, C. and {Dafonte}, C. and {Damerdji}, Y. and {Dapergolas}, A. and {David}, P. and {David}, M. and {De Cat}, P. and {de Felice}, F. and {de Laverny}, P. and {De Luise}, F. and {De March}, R. and {de Martino}, D. and {de Souza}, R. and {Debosscher}, J. and {del Pozo}, E. and {Delbo}, M. and {Delgado}, A. and {Delgado}, H.~E. and {di Marco}, F. and {Di Matteo}, P. and {Diakite}, S. and {Distefano}, E. and {Dolding}, C. and {Dos Anjos}, S. and {Drazinos}, P. and {Dur{\'a}n}, J. and {Dzigan}, Y. and {Ecale}, E. and {Edvardsson}, B. and {Enke}, H. and {Erdmann}, M. and {Escolar}, D. and {Espina}, M. and {Evans}, N.~W. and {Eynard Bontemps}, G. and {Fabre}, C. and {Fabrizio}, M. and {Faigler}, S. and {Falc{\~a}o}, A.~J. and {Farr{\`a}s Casas}, M. and {Faye}, F. and {Federici}, L. and {Fedorets}, G. and {Fern{\'a}ndez-Hern{\'a}ndez}, J. and {Fernique}, P. and {Fienga}, A. and {Figueras}, F. and {Filippi}, F. and {Findeisen}, K. and {Fonti}, A. and {Fouesneau}, M. and {Fraile}, E. and {Fraser}, M. and {Fuchs}, J. and {Furnell}, R. and {Gai}, M. and {Galleti}, S. and {Galluccio}, L. and {Garabato}, D. and {Garc{\'\i}a-Sedano}, F. and {Gar{\'e}}, P. and {Garofalo}, A. and {Garralda}, N. and {Gavras}, P. and {Gerssen}, J. and {Geyer}, R. and {Gilmore}, G. and {Girona}, S. and {Giuffrida}, G. and {Gomes}, M. and {Gonz{\'a}lez-Marcos}, A. and {Gonz{\'a}lez-N{\'u}{\~n}ez}, J. and {Gonz{\'a}lez-Vidal}, J.~J. and {Granvik}, M. and {Guerrier}, A. and {Guillout}, P. and {Guiraud}, J. and {G{\'u}rpide}, A. and {Guti{\'e}rrez-S{\'a}nchez}, R. and {Guy}, L.~P. and {Haigron}, R. and {Hatzidimitriou}, D. and {Haywood}, M. and {Heiter}, U. and {Helmi}, A. and {Hobbs}, D. and {Hofmann}, W. and {Holl}, B. and {Holland}, G. and {Hunt}, J.~A.~S. and {Hypki}, A. and {Icardi}, V. and {Irwin}, M. and {Jevardat de Fombelle}, G. and {Jofr{\'e}}, P. and {Jonker}, P.~G. and {Jorissen}, A. and {Julbe}, F. and {Karampelas}, A. and {Kochoska}, A. and {Kohley}, R. and {Kolenberg}, K. and {Kontizas}, E. and {Koposov}, S.~E. and {Kordopatis}, G. and {Koubsky}, P. and {Kowalczyk}, A. and {Krone-Martins}, A. and {Kudryashova}, M. and {Kull}, I. and {Bachchan}, R.~K. and {Lacoste-Seris}, F. and {Lanza}, A.~F. and {Lavigne}, J. -B. and {Le Poncin-Lafitte}, C. and {Lebreton}, Y. and {Lebzelter}, T. and {Leccia}, S. and {Leclerc}, N. and {Lecoeur-Taibi}, I. and {Lemaitre}, V. and {Lenhardt}, H. and {Leroux}, F. and {Liao}, S. and {Licata}, E. and {Lindstr{\o}m}, H.~E.~P. and {Lister}, T.~A. and {Livanou}, E. and {Lobel}, A. and {L{\"o}ffler}, W. and {L{\'o}pez}, M. and {Lopez-Lozano}, A. and {Lorenz}, D. and {Loureiro}, T. and {MacDonald}, I. and {Magalh{\~a}es Fernandes}, T. and {Managau}, S. and {Mann}, R.~G. and {Mantelet}, G. and {Marchal}, O. and {Marchant}, J.~M. and {Marconi}, M. and {Marie}, J. and {Marinoni}, S. and {Marrese}, P.~M. and {Marschalk{\'o}}, G. and {Marshall}, D.~J. and {Mart{\'\i}n-Fleitas}, J.~M. and {Martino}, M. and {Mary}, N. and {Matijevi{\v{c}}}, G. and {Mazeh}, T. and {McMillan}, P.~J. and {Messina}, S. and {Mestre}, A. and {Michalik}, D. and {Millar}, N.~R. and {Miranda}, B.~M.~H. and {Molina}, D. and {Molinaro}, R. and {Molinaro}, M. and {Moln{\'a}r}, L. and {Moniez}, M. and {Montegriffo}, P. and {Monteiro}, D. and {Mor}, R. and {Mora}, A. and {Morbidelli}, R. and {Morel}, T. and {Morgenthaler}, S. and {Morley}, T. and {Morris}, D. and {Mulone}, A.~F. and {Muraveva}, T. and {Musella}, I. and {Narbonne}, J. and {Nelemans}, G. and {Nicastro}, L. and {Noval}, L. and {Ord{\'e}novic}, C. and {Ordieres-Mer{\'e}}, J. and {Osborne}, P. and {Pagani}, C. and {Pagano}, I. and {Pailler}, F. and {Palacin}, H. and {Palaversa}, L. and {Parsons}, P. and {Paulsen}, T. and {Pecoraro}, M. and {Pedrosa}, R. and {Pentik{\"a}inen}, H. and {Pereira}, J. and {Pichon}, B. and {Piersimoni}, A.~M. and {Pineau}, F. -X. and {Plachy}, E. and {Plum}, G. and {Poujoulet}, E. and {Pr{\v{s}}a}, A. and {Pulone}, L. and {Ragaini}, S. and {Rago}, S. and {Rambaux}, N. and {Ramos-Lerate}, M. and {Ranalli}, P. and {Rauw}, G. and {Read}, A. and {Regibo}, S. and {Renk}, F. and {Reyl{\'e}}, C. and {Ribeiro}, R.~A. and {Rimoldini}, L. and {Ripepi}, V. and {Riva}, A. and {Rixon}, G. and {Roelens}, M. and {Romero-G{\'o}mez}, M. and {Rowell}, N. and {Royer}, F. and {Rudolph}, A. and {Ruiz-Dern}, L. and {Sadowski}, G. and {Sagrist{\`a} Sell{\'e}s}, T. and {Sahlmann}, J. and {Salgado}, J. and {Salguero}, E. and {Sarasso}, M. and {Savietto}, H. and {Schnorhk}, A. and {Schultheis}, M. and {Sciacca}, E. and {Segol}, M. and {Segovia}, J.~C. and {Segransan}, D. and {Serpell}, E. and {Shih}, I. -C. and {Smareglia}, R. and {Smart}, R.~L. and {Smith}, C. and {Solano}, E. and {Solitro}, F. and {Sordo}, R. and {Soria Nieto}, S. and {Souchay}, J. and {Spagna}, A. and {Spoto}, F. and {Stampa}, U. and {Steele}, I.~A. and {Steidelm{\"u}ller}, H. and {Stephenson}, C.~A. and {Stoev}, H. and {Suess}, F.~F. and {S{\"u}veges}, M. and {Surdej}, J. and {Szabados}, L. and {Szegedi-Elek}, E. and {Tapiador}, D. and {Taris}, F. and {Tauran}, G. and {Taylor}, M.~B. and {Teixeira}, R. and {Terrett}, D. and {Tingley}, B. and {Trager}, S.~C. and {Turon}, C. and {Ulla}, A. and {Utrilla}, E. and {Valentini}, G. and {van Elteren}, A. and {Van Hemelryck}, E. and {van Leeuwen}, M. and {Varadi}, M. and {Vecchiato}, A. and {Veljanoski}, J. and {Via}, T. and {Vicente}, D. and {Vogt}, S. and {Voss}, H. and {Votruba}, V. and {Voutsinas}, S. and {Walmsley}, G. and {Weiler}, M. and {Weingrill}, K. and {Werner}, D. and {Wevers}, T. and {Whitehead}, G. and {Wyrzykowski}, {\L}. and {Yoldas}, A. and {{\v{Z}}erjal}, M. and {Zucker}, S. and {Zurbach}, C. and {Zwitter}, T. and {Alecu}, A. and {Allen}, M. and {Allende Prieto}, C. and {Amorim}, A. and {Anglada-Escud{\'e}}, G. and {Arsenijevic}, V. and {Azaz}, S. and {Balm}, P. and {Beck}, M. and {Bernstein}, H. -H. and {Bigot}, L. and {Bijaoui}, A. and {Blasco}, C. and {Bonfigli}, M. and {Bono}, G. and {Boudreault}, S. and {Bressan}, A. and {Brown}, S. and {Brunet}, P. -M. and {Bunclark}, P. and {Buonanno}, R. and {Butkevich}, A.~G. and {Carret}, C. and {Carrion}, C. and {Chemin}, L. and {Ch{\'e}reau}, F. and {Corcione}, L. and {Darmigny}, E. and {de Boer}, K.~S. and {de Teodoro}, P. and {de Zeeuw}, P.~T. and {Delle Luche}, C. and {Domingues}, C.~D. and {Dubath}, P. and {Fodor}, F. and {Fr{\'e}zouls}, B. and {Fries}, A. and {Fustes}, D. and {Fyfe}, D. and {Gallardo}, E. and {Gallegos}, J. and {Gardiol}, D. and {Gebran}, M. and {Gomboc}, A. and {G{\'o}mez}, A. and {Grux}, E. and {Gueguen}, A. and {Heyrovsky}, A. and {Hoar}, J. and {Iannicola}, G. and {Isasi Parache}, Y. and {Janotto}, A. -M. and {Joliet}, E. and {Jonckheere}, A. and {Keil}, R. and {Kim}, D. -W. and {Klagyivik}, P. and {Klar}, J. and {Knude}, J. and {Kochukhov}, O. and {Kolka}, I. and {Kos}, J. and {Kutka}, A. and {Lainey}, V. and {LeBouquin}, D. and {Liu}, C. and {Loreggia}, D. and {Makarov}, V.~V. and {Marseille}, M.~G. and {Martayan}, C. and {Martinez-Rubi}, O. and {Massart}, B. and {Meynadier}, F. and {Mignot}, S. and {Munari}, U. and {Nguyen}, A. -T. and {Nordlander}, T. and {Ocvirk}, P. and {O'Flaherty}, K.~S. and {Olias Sanz}, A. and {Ortiz}, P. and {Osorio}, J. and {Oszkiewicz}, D. and {Ouzounis}, A. and {Palmer}, M. and {Park}, P. and {Pasquato}, E. and {Peltzer}, C. and {Peralta}, J. and {P{\'e}turaud}, F. and {Pieniluoma}, T. and {Pigozzi}, E. and {Poels}, J. and {Prat}, G. and {Prod'homme}, T. and {Raison}, F. and {Rebordao}, J.~M. and {Risquez}, D. and {Rocca-Volmerange}, B. and {Rosen}, S. and {Ruiz-Fuertes}, M.~I. and {Russo}, F. and {Sembay}, S. and {Serraller Vizcaino}, I. and {Short}, A. and {Siebert}, A. and {Silva}, H. and {Sinachopoulos}, D. and {Slezak}, E. and {Soffel}, M. and {Sosnowska}, D. and {Strai{\v{z}}ys}, V. and {ter Linden}, M. and {Terrell}, D. and {Theil}, S. and {Tiede}, C. and {Troisi}, L. and {Tsalmantza}, P. and {Tur}, D. and {Vaccari}, M. and {Vachier}, F. and {Valles}, P. and {Van Hamme}, W. and {Veltz}, L. and {Virtanen}, J. and {Wallut}, J. -M. and {Wichmann}, R. and {Wilkinson}, M.~I. and {Ziaeepour}, H. and {Zschocke}, S.},
        title = "{The Gaia mission}",
      journal = {\aap},
         year = 2016,
        month = nov,
       volume = {595},
          eid = {A1},
        pages = {A1},
          doi = {10.1051/0004-6361/201629272},
archivePrefix = {arXiv},
       eprint = {1609.04153},
 primaryClass = {astro-ph.IM},
       adsurl = {https://ui.adsabs.harvard.edu/abs/2016A&A...595A...1G}
}

@ARTICLE{Gaia_2018,
       author = {{Gaia Collaboration} and {Brown}, A.~G.~A. and {Vallenari}, A. and {Prusti}, T. and {de Bruijne}, J.~H.~J. and {Babusiaux}, C. and {Bailer-Jones}, C.~A.~L. and {Biermann}, M. and {Evans}, D.~W. and {Eyer}, L. and {Jansen}, F. and {Jordi}, C. and {Klioner}, S.~A. and {Lammers}, U. and {Lindegren}, L. and {Luri}, X. and {Mignard}, F. and {Panem}, C. and {Pourbaix}, D. and {Randich}, S. and {Sartoretti}, P. and {Siddiqui}, H.~I. and {Soubiran}, C. and {van Leeuwen}, F. and {Walton}, N.~A. and {Arenou}, F. and {Bastian}, U. and {Cropper}, M. and {Drimmel}, R. and {Katz}, D. and {Lattanzi}, M.~G. and {Bakker}, J. and {Cacciari}, C. and {Casta{\~n}eda}, J. and {Chaoul}, L. and {Cheek}, N. and {De Angeli}, F. and {Fabricius}, C. and {Guerra}, R. and {Holl}, B. and {Masana}, E. and {Messineo}, R. and {Mowlavi}, N. and {Nienartowicz}, K. and {Panuzzo}, P. and {Portell}, J. and {Riello}, M. and {Seabroke}, G.~M. and {Tanga}, P. and {Th{\'e}venin}, F. and {Gracia-Abril}, G. and {Comoretto}, G. and {Garcia-Reinaldos}, M. and {Teyssier}, D. and {Altmann}, M. and {Andrae}, R. and {Audard}, M. and {Bellas-Velidis}, I. and {Benson}, K. and {Berthier}, J. and {Blomme}, R. and {Burgess}, P. and {Busso}, G. and {Carry}, B. and {Cellino}, A. and {Clementini}, G. and {Clotet}, M. and {Creevey}, O. and {Davidson}, M. and {De Ridder}, J. and {Delchambre}, L. and {Dell'Oro}, A. and {Ducourant}, C. and {Fern{\'a}ndez-Hern{\'a}ndez}, J. and {Fouesneau}, M. and {Fr{\'e}mat}, Y. and {Galluccio}, L. and {Garc{\'\i}a-Torres}, M. and {Gonz{\'a}lez-N{\'u}{\~n}ez}, J. and {Gonz{\'a}lez-Vidal}, J.~J. and {Gosset}, E. and {Guy}, L.~P. and {Halbwachs}, J. -L. and {Hambly}, N.~C. and {Harrison}, D.~L. and {Hern{\'a}ndez}, J. and {Hestroffer}, D. and {Hodgkin}, S.~T. and {Hutton}, A. and {Jasniewicz}, G. and {Jean-Antoine-Piccolo}, A. and {Jordan}, S. and {Korn}, A.~J. and {Krone-Martins}, A. and {Lanzafame}, A.~C. and {Lebzelter}, T. and {L{\"o}ffler}, W. and {Manteiga}, M. and {Marrese}, P.~M. and {Mart{\'\i}n-Fleitas}, J.~M. and {Moitinho}, A. and {Mora}, A. and {Muinonen}, K. and {Osinde}, J. and {Pancino}, E. and {Pauwels}, T. and {Petit}, J. -M. and {Recio-Blanco}, A. and {Richards}, P.~J. and {Rimoldini}, L. and {Robin}, A.~C. and {Sarro}, L.~M. and {Siopis}, C. and {Smith}, M. and {Sozzetti}, A. and {S{\"u}veges}, M. and {Torra}, J. and {van Reeven}, W. and {Abbas}, U. and {Abreu Aramburu}, A. and {Accart}, S. and {Aerts}, C. and {Altavilla}, G. and {{\'A}lvarez}, M.~A. and {Alvarez}, R. and {Alves}, J. and {Anderson}, R.~I. and {Andrei}, A.~H. and {Anglada Varela}, E. and {Antiche}, E. and {Antoja}, T. and {Arcay}, B. and {Astraatmadja}, T.~L. and {Bach}, N. and {Baker}, S.~G. and {Balaguer-N{\'u}{\~n}ez}, L. and {Balm}, P. and {Barache}, C. and {Barata}, C. and {Barbato}, D. and {Barblan}, F. and {Barklem}, P.~S. and {Barrado}, D. and {Barros}, M. and {Barstow}, M.~A. and {Bartholom{\'e} Mu{\~n}oz}, S. and {Bassilana}, J. -L. and {Becciani}, U. and {Bellazzini}, M. and {Berihuete}, A. and {Bertone}, S. and {Bianchi}, L. and {Bienaym{\'e}}, O. and {Blanco-Cuaresma}, S. and {Boch}, T. and {Boeche}, C. and {Bombrun}, A. and {Borrachero}, R. and {Bossini}, D. and {Bouquillon}, S. and {Bourda}, G. and {Bragaglia}, A. and {Bramante}, L. and {Breddels}, M.~A. and {Bressan}, A. and {Brouillet}, N. and {Br{\"u}semeister}, T. and {Brugaletta}, E. and {Bucciarelli}, B. and {Burlacu}, A. and {Busonero}, D. and {Butkevich}, A.~G. and {Buzzi}, R. and {Caffau}, E. and {Cancelliere}, R. and {Cannizzaro}, G. and {Cantat-Gaudin}, T. and {Carballo}, R. and {Carlucci}, T. and {Carrasco}, J.~M. and {Casamiquela}, L. and {Castellani}, M. and {Castro-Ginard}, A. and {Charlot}, P. and {Chemin}, L. and {Chiavassa}, A. and {Cocozza}, G. and {Costigan}, G. and {Cowell}, S. and {Crifo}, F. and {Crosta}, M. and {Crowley}, C. and {Cuypers}, J. and {Dafonte}, C. and {Damerdji}, Y. and {Dapergolas}, A. and {David}, P. and {David}, M. and {de Laverny}, P. and {De Luise}, F. and {De March}, R. and {de Martino}, D. and {de Souza}, R. and {de Torres}, A. and {Debosscher}, J. and {del Pozo}, E. and {Delbo}, M. and {Delgado}, A. and {Delgado}, H.~E. and {Di Matteo}, P. and {Diakite}, S. and {Diener}, C. and {Distefano}, E. and {Dolding}, C. and {Drazinos}, P. and {Dur{\'a}n}, J. and {Edvardsson}, B. and {Enke}, H. and {Eriksson}, K. and {Esquej}, P. and {Eynard Bontemps}, G. and {Fabre}, C. and {Fabrizio}, M. and {Faigler}, S. and {Falc{\~a}o}, A.~J. and {Farr{\`a}s Casas}, M. and {Federici}, L. and {Fedorets}, G. and {Fernique}, P. and {Figueras}, F. and {Filippi}, F. and {Findeisen}, K. and {Fonti}, A. and {Fraile}, E. and {Fraser}, M. and {Fr{\'e}zouls}, B. and {Gai}, M. and {Galleti}, S. and {Garabato}, D. and {Garc{\'\i}a-Sedano}, F. and {Garofalo}, A. and {Garralda}, N. and {Gavel}, A. and {Gavras}, P. and {Gerssen}, J. and {Geyer}, R. and {Giacobbe}, P. and {Gilmore}, G. and {Girona}, S. and {Giuffrida}, G. and {Glass}, F. and {Gomes}, M. and {Granvik}, M. and {Gueguen}, A. and {Guerrier}, A. and {Guiraud}, J. and {Guti{\'e}rrez-S{\'a}nchez}, R. and {Haigron}, R. and {Hatzidimitriou}, D. and {Hauser}, M. and {Haywood}, M. and {Heiter}, U. and {Helmi}, A. and {Heu}, J. and {Hilger}, T. and {Hobbs}, D. and {Hofmann}, W. and {Holland}, G. and {Huckle}, H.~E. and {Hypki}, A. and {Icardi}, V. and {Jan{\ss}en}, K. and {Jevardat de Fombelle}, G. and {Jonker}, P.~G. and {Juh{\'a}sz}, {\'A}. L. and {Julbe}, F. and {Karampelas}, A. and {Kewley}, A. and {Klar}, J. and {Kochoska}, A. and {Kohley}, R. and {Kolenberg}, K. and {Kontizas}, M. and {Kontizas}, E. and {Koposov}, S.~E. and {Kordopatis}, G. and {Kostrzewa-Rutkowska}, Z. and {Koubsky}, P. and {Lambert}, S. and {Lanza}, A.~F. and {Lasne}, Y. and {Lavigne}, J. -B. and {Le Fustec}, Y. and {Le Poncin-Lafitte}, C. and {Lebreton}, Y. and {Leccia}, S. and {Leclerc}, N. and {Lecoeur-Taibi}, I. and {Lenhardt}, H. and {Leroux}, F. and {Liao}, S. and {Licata}, E. and {Lindstr{\o}m}, H.~E.~P. and {Lister}, T.~A. and {Livanou}, E. and {Lobel}, A. and {L{\'o}pez}, M. and {Managau}, S. and {Mann}, R.~G. and {Mantelet}, G. and {Marchal}, O. and {Marchant}, J.~M. and {Marconi}, M. and {Marinoni}, S. and {Marschalk{\'o}}, G. and {Marshall}, D.~J. and {Martino}, M. and {Marton}, G. and {Mary}, N. and {Massari}, D. and {Matijevi{\v{c}}}, G. and {Mazeh}, T. and {McMillan}, P.~J. and {Messina}, S. and {Michalik}, D. and {Millar}, N.~R. and {Molina}, D. and {Molinaro}, R. and {Moln{\'a}r}, L. and {Montegriffo}, P. and {Mor}, R. and {Morbidelli}, R. and {Morel}, T. and {Morris}, D. and {Mulone}, A.~F. and {Muraveva}, T. and {Musella}, I. and {Nelemans}, G. and {Nicastro}, L. and {Noval}, L. and {O'Mullane}, W. and {Ord{\'e}novic}, C. and {Ord{\'o}{\~n}ez-Blanco}, D. and {Osborne}, P. and {Pagani}, C. and {Pagano}, I. and {Pailler}, F. and {Palacin}, H. and {Palaversa}, L. and {Panahi}, A. and {Pawlak}, M. and {Piersimoni}, A.~M. and {Pineau}, F. -X. and {Plachy}, E. and {Plum}, G. and {Poggio}, E. and {Poujoulet}, E. and {Pr{\v{s}}a}, A. and {Pulone}, L. and {Racero}, E. and {Ragaini}, S. and {Rambaux}, N. and {Ramos-Lerate}, M. and {Regibo}, S. and {Reyl{\'e}}, C. and {Riclet}, F. and {Ripepi}, V. and {Riva}, A. and {Rivard}, A. and {Rixon}, G. and {Roegiers}, T. and {Roelens}, M. and {Romero-G{\'o}mez}, M. and {Rowell}, N. and {Royer}, F. and {Ruiz-Dern}, L. and {Sadowski}, G. and {Sagrist{\`a} Sell{\'e}s}, T. and {Sahlmann}, J. and {Salgado}, J. and {Salguero}, E. and {Sanna}, N. and {Santana-Ros}, T. and {Sarasso}, M. and {Savietto}, H. and {Schultheis}, M. and {Sciacca}, E. and {Segol}, M. and {Segovia}, J.~C. and {S{\'e}gransan}, D. and {Shih}, I. -C. and {Siltala}, L. and {Silva}, A.~F. and {Smart}, R.~L. and {Smith}, K.~W. and {Solano}, E. and {Solitro}, F. and {Sordo}, R. and {Soria Nieto}, S. and {Souchay}, J. and {Spagna}, A. and {Spoto}, F. and {Stampa}, U. and {Steele}, I.~A. and {Steidelm{\"u}ller}, H. and {Stephenson}, C.~A. and {Stoev}, H. and {Suess}, F.~F. and {Surdej}, J. and {Szabados}, L. and {Szegedi-Elek}, E. and {Tapiador}, D. and {Taris}, F. and {Tauran}, G. and {Taylor}, M.~B. and {Teixeira}, R. and {Terrett}, D. and {Teyssandier}, P. and {Thuillot}, W. and {Titarenko}, A. and {Torra Clotet}, F. and {Turon}, C. and {Ulla}, A. and {Utrilla}, E. and {Uzzi}, S. and {Vaillant}, M. and {Valentini}, G. and {Valette}, V. and {van Elteren}, A. and {Van Hemelryck}, E. and {van Leeuwen}, M. and {Vaschetto}, M. and {Vecchiato}, A. and {Veljanoski}, J. and {Viala}, Y. and {Vicente}, D. and {Vogt}, S. and {von Essen}, C. and {Voss}, H. and {Votruba}, V. and {Voutsinas}, S. and {Walmsley}, G. and {Weiler}, M. and {Wertz}, O. and {Wevers}, T. and {Wyrzykowski}, {\L}. and {Yoldas}, A. and {{\v{Z}}erjal}, M. and {Ziaeepour}, H. and {Zorec}, J. and {Zschocke}, S. and {Zucker}, S. and {Zurbach}, C. and {Zwitter}, T.},
        title = "{Gaia Data Release 2. Summary of the contents and survey properties}",
      journal = {\aap},
         year = 2018,
        month = aug,
       volume = {616},
          eid = {A1},
        pages = {A1},
          doi = {10.1051/0004-6361/201833051},
archivePrefix = {arXiv},
       eprint = {1804.09365},
 primaryClass = {astro-ph.GA},
       adsurl = {https://ui.adsabs.harvard.edu/abs/2018A&A...616A...1G}
}

@ARTICLE{Gaia_2021,
       author = {{Gaia Collaboration} and {Brown}, A.~G.~A. and {Vallenari}, A. and {Prusti}, T. and {de Bruijne}, J.~H.~J. and {Babusiaux}, C. and {Biermann}, M. and {Creevey}, O.~L. and {Evans}, D.~W. and {Eyer}, L. and {Hutton}, A. and {Jansen}, F. and {Jordi}, C. and {Klioner}, S.~A. and {Lammers}, U. and {Lindegren}, L. and {Luri}, X. and {Mignard}, F. and {Panem}, C. and {Pourbaix}, D. and {Randich}, S. and {Sartoretti}, P. and {Soubiran}, C. and {Walton}, N.~A. and {Arenou}, F. and {Bailer-Jones}, C.~A.~L. and {Bastian}, U. and {Cropper}, M. and {Drimmel}, R. and {Katz}, D. and {Lattanzi}, M.~G. and {van Leeuwen}, F. and {Bakker}, J. and {Cacciari}, C. and {Casta{\~n}eda}, J. and {De Angeli}, F. and {Ducourant}, C. and {Fabricius}, C. and {Fouesneau}, M. and {Fr{\'e}mat}, Y. and {Guerra}, R. and {Guerrier}, A. and {Guiraud}, J. and {Jean-Antoine Piccolo}, A. and {Masana}, E. and {Messineo}, R. and {Mowlavi}, N. and {Nicolas}, C. and {Nienartowicz}, K. and {Pailler}, F. and {Panuzzo}, P. and {Riclet}, F. and {Roux}, W. and {Seabroke}, G.~M. and {Sordo}, R. and {Tanga}, P. and {Th{\'e}venin}, F. and {Gracia-Abril}, G. and {Portell}, J. and {Teyssier}, D. and {Altmann}, M. and {Andrae}, R. and {Bellas-Velidis}, I. and {Benson}, K. and {Berthier}, J. and {Blomme}, R. and {Brugaletta}, E. and {Burgess}, P.~W. and {Busso}, G. and {Carry}, B. and {Cellino}, A. and {Cheek}, N. and {Clementini}, G. and {Damerdji}, Y. and {Davidson}, M. and {Delchambre}, L. and {Dell'Oro}, A. and {Fern{\'a}ndez-Hern{\'a}ndez}, J. and {Galluccio}, L. and {Garc{\'\i}a-Lario}, P. and {Garcia-Reinaldos}, M. and {Gonz{\'a}lez-N{\'u}{\~n}ez}, J. and {Gosset}, E. and {Haigron}, R. and {Halbwachs}, J. -L. and {Hambly}, N.~C. and {Harrison}, D.~L. and {Hatzidimitriou}, D. and {Heiter}, U. and {Hern{\'a}ndez}, J. and {Hestroffer}, D. and {Hodgkin}, S.~T. and {Holl}, B. and {Jan{\ss}en}, K. and {Jevardat de Fombelle}, G. and {Jordan}, S. and {Krone-Martins}, A. and {Lanzafame}, A.~C. and {L{\"o}ffler}, W. and {Lorca}, A. and {Manteiga}, M. and {Marchal}, O. and {Marrese}, P.~M. and {Moitinho}, A. and {Mora}, A. and {Muinonen}, K. and {Osborne}, P. and {Pancino}, E. and {Pauwels}, T. and {Petit}, J. -M. and {Recio-Blanco}, A. and {Richards}, P.~J. and {Riello}, M. and {Rimoldini}, L. and {Robin}, A.~C. and {Roegiers}, T. and {Rybizki}, J. and {Sarro}, L.~M. and {Siopis}, C. and {Smith}, M. and {Sozzetti}, A. and {Ulla}, A. and {Utrilla}, E. and {van Leeuwen}, M. and {van Reeven}, W. and {Abbas}, U. and {Abreu Aramburu}, A. and {Accart}, S. and {Aerts}, C. and {Aguado}, J.~J. and {Ajaj}, M. and {Altavilla}, G. and {{\'A}lvarez}, M.~A. and {{\'A}lvarez Cid-Fuentes}, J. and {Alves}, J. and {Anderson}, R.~I. and {Anglada Varela}, E. and {Antoja}, T. and {Audard}, M. and {Baines}, D. and {Baker}, S.~G. and {Balaguer-N{\'u}{\~n}ez}, L. and {Balbinot}, E. and {Balog}, Z. and {Barache}, C. and {Barbato}, D. and {Barros}, M. and {Barstow}, M.~A. and {Bartolom{\'e}}, S. and {Bassilana}, J. -L. and {Bauchet}, N. and {Baudesson-Stella}, A. and {Becciani}, U. and {Bellazzini}, M. and {Bernet}, M. and {Bertone}, S. and {Bianchi}, L. and {Blanco-Cuaresma}, S. and {Boch}, T. and {Bombrun}, A. and {Bossini}, D. and {Bouquillon}, S. and {Bragaglia}, A. and {Bramante}, L. and {Breedt}, E. and {Bressan}, A. and {Brouillet}, N. and {Bucciarelli}, B. and {Burlacu}, A. and {Busonero}, D. and {Butkevich}, A.~G. and {Buzzi}, R. and {Caffau}, E. and {Cancelliere}, R. and {C{\'a}novas}, H. and {Cantat-Gaudin}, T. and {Carballo}, R. and {Carlucci}, T. and {Carnerero}, M.~I. and {Carrasco}, J.~M. and {Casamiquela}, L. and {Castellani}, M. and {Castro-Ginard}, A. and {Castro Sampol}, P. and {Chaoul}, L. and {Charlot}, P. and {Chemin}, L. and {Chiavassa}, A. and {Cioni}, M. -R.~L. and {Comoretto}, G. and {Cooper}, W.~J. and {Cornez}, T. and {Cowell}, S. and {Crifo}, F. and {Crosta}, M. and {Crowley}, C. and {Dafonte}, C. and {Dapergolas}, A. and {David}, M. and {David}, P. and {de Laverny}, P. and {De Luise}, F. and {De March}, R. and {De Ridder}, J. and {de Souza}, R. and {de Teodoro}, P. and {de Torres}, A. and {del Peloso}, E.~F. and {del Pozo}, E. and {Delbo}, M. and {Delgado}, A. and {Delgado}, H.~E. and {Delisle}, J. -B. and {Di Matteo}, P. and {Diakite}, S. and {Diener}, C. and {Distefano}, E. and {Dolding}, C. and {Eappachen}, D. and {Edvardsson}, B. and {Enke}, H. and {Esquej}, P. and {Fabre}, C. and {Fabrizio}, M. and {Faigler}, S. and {Fedorets}, G. and {Fernique}, P. and {Fienga}, A. and {Figueras}, F. and {Fouron}, C. and {Fragkoudi}, F. and {Fraile}, E. and {Franke}, F. and {Gai}, M. and {Garabato}, D. and {Garcia-Gutierrez}, A. and {Garc{\'\i}a-Torres}, M. and {Garofalo}, A. and {Gavras}, P. and {Gerlach}, E. and {Geyer}, R. and {Giacobbe}, P. and {Gilmore}, G. and {Girona}, S. and {Giuffrida}, G. and {Gomel}, R. and {Gomez}, A. and {Gonzalez-Santamaria}, I. and {Gonz{\'a}lez-Vidal}, J.~J. and {Granvik}, M. and {Guti{\'e}rrez-S{\'a}nchez}, R. and {Guy}, L.~P. and {Hauser}, M. and {Haywood}, M. and {Helmi}, A. and {Hidalgo}, S.~L. and {Hilger}, T. and {H{\l}adczuk}, N. and {Hobbs}, D. and {Holland}, G. and {Huckle}, H.~E. and {Jasniewicz}, G. and {Jonker}, P.~G. and {Juaristi Campillo}, J. and {Julbe}, F. and {Karbevska}, L. and {Kervella}, P. and {Khanna}, S. and {Kochoska}, A. and {Kontizas}, M. and {Kordopatis}, G. and {Korn}, A.~J. and {Kostrzewa-Rutkowska}, Z. and {Kruszy{\'n}ska}, K. and {Lambert}, S. and {Lanza}, A.~F. and {Lasne}, Y. and {Le Campion}, J. -F. and {Le Fustec}, Y. and {Lebreton}, Y. and {Lebzelter}, T. and {Leccia}, S. and {Leclerc}, N. and {Lecoeur-Taibi}, I. and {Liao}, S. and {Licata}, E. and {Lindstr{\o}m}, E.~P. and {Lister}, T.~A. and {Livanou}, E. and {Lobel}, A. and {Madrero Pardo}, P. and {Managau}, S. and {Mann}, R.~G. and {Marchant}, J.~M. and {Marconi}, M. and {Marcos Santos}, M.~M.~S. and {Marinoni}, S. and {Marocco}, F. and {Marshall}, D.~J. and {Martin Polo}, L. and {Mart{\'\i}n-Fleitas}, J.~M. and {Masip}, A. and {Massari}, D. and {Mastrobuono-Battisti}, A. and {Mazeh}, T. and {McMillan}, P.~J. and {Messina}, S. and {Michalik}, D. and {Millar}, N.~R. and {Mints}, A. and {Molina}, D. and {Molinaro}, R. and {Moln{\'a}r}, L. and {Montegriffo}, P. and {Mor}, R. and {Morbidelli}, R. and {Morel}, T. and {Morris}, D. and {Mulone}, A.~F. and {Munoz}, D. and {Muraveva}, T. and {Murphy}, C.~P. and {Musella}, I. and {Noval}, L. and {Ord{\'e}novic}, C. and {Orr{\`u}}, G. and {Osinde}, J. and {Pagani}, C. and {Pagano}, I. and {Palaversa}, L. and {Palicio}, P.~A. and {Panahi}, A. and {Pawlak}, M. and {Pe{\~n}alosa Esteller}, X. and {Penttil{\"a}}, A. and {Piersimoni}, A.~M. and {Pineau}, F. -X. and {Plachy}, E. and {Plum}, G. and {Poggio}, E. and {Poretti}, E. and {Poujoulet}, E. and {Pr{\v{s}}a}, A. and {Pulone}, L. and {Racero}, E. and {Ragaini}, S. and {Rainer}, M. and {Raiteri}, C.~M. and {Rambaux}, N. and {Ramos}, P. and {Ramos-Lerate}, M. and {Re Fiorentin}, P. and {Regibo}, S. and {Reyl{\'e}}, C. and {Ripepi}, V. and {Riva}, A. and {Rixon}, G. and {Robichon}, N. and {Robin}, C. and {Roelens}, M. and {Rohrbasser}, L. and {Romero-G{\'o}mez}, M. and {Rowell}, N. and {Royer}, F. and {Rybicki}, K.~A. and {Sadowski}, G. and {Sagrist{\`a} Sell{\'e}s}, A. and {Sahlmann}, J. and {Salgado}, J. and {Salguero}, E. and {Samaras}, N. and {Sanchez Gimenez}, V. and {Sanna}, N. and {Santove{\~n}a}, R. and {Sarasso}, M. and {Schultheis}, M. and {Sciacca}, E. and {Segol}, M. and {Segovia}, J.~C. and {S{\'e}gransan}, D. and {Semeux}, D. and {Shahaf}, S. and {Siddiqui}, H.~I. and {Siebert}, A. and {Siltala}, L. and {Slezak}, E. and {Smart}, R.~L. and {Solano}, E. and {Solitro}, F. and {Souami}, D. and {Souchay}, J. and {Spagna}, A. and {Spoto}, F. and {Steele}, I.~A. and {Steidelm{\"u}ller}, H. and {Stephenson}, C.~A. and {S{\"u}veges}, M. and {Szabados}, L. and {Szegedi-Elek}, E. and {Taris}, F. and {Tauran}, G. and {Taylor}, M.~B. and {Teixeira}, R. and {Thuillot}, W. and {Tonello}, N. and {Torra}, F. and {Torra}, J. and {Turon}, C. and {Unger}, N. and {Vaillant}, M. and {van Dillen}, E. and {Vanel}, O. and {Vecchiato}, A. and {Viala}, Y. and {Vicente}, D. and {Voutsinas}, S. and {Weiler}, M. and {Wevers}, T. and {Wyrzykowski}, {\L}. and {Yoldas}, A. and {Yvard}, P. and {Zhao}, H. and {Zorec}, J. and {Zucker}, S. and {Zurbach}, C. and {Zwitter}, T.},
        title = "{Gaia Early Data Release 3. Summary of the contents and survey properties}",
      journal = {\aap},
         year = 2021,
        month = may,
       volume = {649},
          eid = {A1},
        pages = {A1},
          doi = {10.1051/0004-6361/202039657},
archivePrefix = {arXiv},
       eprint = {2012.01533},
 primaryClass = {astro-ph.GA},
       adsurl = {https://ui.adsabs.harvard.edu/abs/2021A&A...649A...1G}
}

@ARTICLE{Shibata_2011,
       author = {{Shibata}, Kazunari and {Magara}, Tetsuya},
        title = "{Solar Flares: Magnetohydrodynamic Processes}",
      journal = {Living Reviews in Solar Physics},
         year = 2011,
        month = dec,
       volume = {8},
       number = {1},
          eid = {6},
        pages = {6},
          doi = {10.12942/lrsp-2011-6},
       adsurl = {https://ui.adsabs.harvard.edu/abs/2011LRSP....8....6S}
}

@ARTICLE{Franciosini_2001,
       author = {{Franciosini}, E. and {Pallavicini}, R. and {Tagliaferri}, G.},
        title = "{BeppoSAX observation of a large long-duration X-ray flare from UX Arietis}",
      journal = {\aap},
         year = 2001,
        month = aug,
       volume = {375},
        pages = {196-204},
          doi = {10.1051/0004-6361:20010830},
       adsurl = {https://ui.adsabs.harvard.edu/abs/2001A&A...375..196F}
}

@software{heasoft_2014,
       author = {{HEASARC}},
        title = "{HEAsoft: Unified Release of FTOOLS and XANADU}",
 howpublished = {Astrophysics Source Code Library, record ascl:1408.004},
         year = 2014,
        month = aug,
          eid = {ascl:1408.004},
       adsurl = {https://ui.adsabs.harvard.edu/abs/2014ascl.soft08004N}
}

@ARTICLE{Anders_1989,
       author = {{Anders}, E. and {Grevesse}, N.},
        title = "{Abundances of the elements: Meteoritic and solar}",
      journal = {\gca},
         year = 1989,
        month = jan,
       volume = {53},
       number = {1},
        pages = {197-214},
          doi = {10.1016/0016-7037(89)90286-X},
       adsurl = {https://ui.adsabs.harvard.edu/abs/1989GeCoA..53..197A}
}

@ARTICLE{Inoue_2025,
       author = {{Inoue}, Shun and {Enoto}, Teruaki and {Notsu}, Yuta and {Uchida}, Hiroyuki and {Iwakiri}, Wataru Buz and {Namekata}, Kosuke and {Gendreau}, Keith},
        title = "{Systematic NICER study of the low-ionized Fe K{\ensuremath{\alpha}} line on RS Canum Venaticorum-type stars}",
      journal = {\mnras},
         year = "2025",
        month = aug,
       volume = {541},
       number = {2},
        pages = {1403-1418},
          doi = {10.1093/mnras/staf1048},
archivePrefix = {arXiv},
       eprint = {2506.21040},
 primaryClass = {astro-ph.SR},
       adsurl = {https://ui.adsabs.harvard.edu/abs/2025MNRAS.541.1403I}
}

@ARTICLE{Osten_1999,
       author = {{Osten}, Rachel A. and {Brown}, Alexander},
        title = "{Extreme Ultraviolet Explorer Photometry of RS Canum Venaticorum Systems: Four Flaring Megaseconds}",
      journal = {\apj},
         year = 1999,
        month = apr,
       volume = {515},
       number = {2},
        pages = {746-761},
          doi = {10.1086/307034},
       adsurl = {https://ui.adsabs.harvard.edu/abs/1999ApJ...515..746O}
}
\bibliographystyle{apj}

\end{document}